\documentclass[12pt]{article}
\pdfoutput=1  
\usepackage{graphicx}
\usepackage[tbtags]{amsmath}
\usepackage{amssymb}
\usepackage{amsthm}
\usepackage{array}
\usepackage{xy}
\usepackage{slashed}
\usepackage[margin=15mm]{geometry}
\usepackage{tikz}
\usepackage{slashbox}
\usepackage{wasysym}

\usepackage{amssymb}
\usepackage{eucal}
\usepackage{amssymb}
\usepackage{mathrsfs}

\usepackage{adjustbox}

\usepackage{color}

\usepackage{verbatim}
\usepackage{booktabs}
\usepackage{multirow}

\usepackage[iso-8859-7]{inputenc}

\def\<{\left\langle}
\def\>{\right\rangle}

\begin{document} 
	
\begin{titlepage}
	\thispagestyle{empty}
	\begin{flushright}

	\end{flushright}

	\vspace{35pt}
	
	\begin{center}
	    { \Large{\bf Probing the BSM physics with CMB precision cosmology: an application to supersymmetry}}
		
		\vspace{50pt}

		{Ioannis Dalianis$^{1}$ and Yuki Watanabe$^{2}$} \makeatletter{\renewcommand*{\@makefnmark}{}
\footnotetext{E-mail addresses: dalianis@mail.ntua.gr,\, yuki.watanabe@nat.gunma-ct.ac.jp}\makeatother} 

		\vspace{25pt}

		{
			$^1${\it  Physics Division, National Technical University of Athens, 
			\\ \it 15780 Zografou Campus, Athens, Greece} 

		\vspace{15pt}

			$^2${\it  Department of Physics, National Institute of Technology, Gunma College, Gunma 371-8530,
Japan}

		}

		\vspace{40pt}

{ABSTRACT}
	\end{center}

\vspace{10pt} 
\noindent 
The cosmic history before the BBN is highly determined by the physics that operates beyond the Standard Model (BSM) of particle physics and it is poorly constrained observationally. 
Ongoing and future precision measurements of the CMB observables can provide us with significant information about the pre-BBN era 
and hence possibly test the cosmological predictions of different BSM scenarios. 
Supersymmetry is a particularly motivated BSM theory and it is often the case that 
different superymmetry breaking schemes require different cosmic histories with specific reheating temperatures or low entropy production in order to be cosmologically viable.
In this paper we quantify the effects of the possible alternative cosmic histories on the $n_s$ and $r$ CMB observables assuming a generic non-thermal stage after cosmic inflation. 
We analyze TeV and especially multi-TeV supersymmetry breaking schemes 
assuming the neutralino and gravitino dark matter scenarios. 
We complement our analysis considering the Starobinsky $R^2$ inflation model to exemplify the improved CMB predictions that a unified description of the early universe cosmic evolution yields.  
Our analysis underlines the importance of the CMB precision measurements that can be viewed, to some extend,  as complementary to the laboratory experimental searches for supersymmetry or other BSM theories.

\bigskip

\end{titlepage}

\baselineskip 6 mm

\tableofcontents

\section{Introduction}

The cosmic evolution before the Big Bang Nucleosynthesis (BBN) and after inflation is much unknown. To date there are no direct observational probes that can constrain this very early universe period, that can be called {\itshape dark pre-BBN period}. 
On the other hand,  inflation that takes place at energy scales much higher than the BBN gives concrete predictions thanks to the presence of the quasi-de Sitter horizon. It is actually the dark pre-BBN cosmic phase that introduces an uncertainty at the inflationary predictions parametrized by the number of e-folds $N_*$. This uncertainty could be minimized if the physics that operates beyond the Standard Model of particle physics (BSM) was known.  Indeed, different BSM scenarios often imply a different cosmic evolution in order to satisfy the BBN predictions and the observed dark matter abundance $\Omega_\text{DM}h^2=0.12$ \cite{Ade:2015xua, Ade:2015lrj}.

The fact that  the $N_*$ is modified by the details of the dark pre-BBN stage \cite{Liddle:2003as} motivate us to investigate this small but non-zero residual dependence of the inflationary predictions on the tentative  BSM physics. 
In most of the inflationary models, a precise measurement of the spectral index $n_s(N_*)$ and tensor-to-scalar ratio $r(N_*)$ value accounts for an indirect measure of the reheating temperature of the universe \cite{Martin:2010kz, Mielczarek:2010ag, Bezrukov:2011gp, Martin:2014nya, Munoz:2014eqa, Dai:2014jja, Gong:2015qha, Cook:2015vqa, Rehagen:2015zma, Drewes:2015coa} and hence one could in principle examine the cosmology of theories beyond the Standard Model of particle physics as well as non-trivial extensions of the Einstein gravity \cite{Dalianis:2016wpu}. 
From the inflation phenomenology point of view, for a given concrete BSM scenario a predictive inflationary model can be spotted on the ($n_s, r$) plane, whereas from the particle physicist   point of view, for a given predictable inflationary scenario the precise measurement of the ($n_s, r$) observables is a measurement of the BSM effects on the cosmic evolution. 
In other words, we can say that the ($n_s, r$) precision measurements provide us with a {\itshape cosmic selection criterion} for the assumed BSM physics. Planck collaboration has constrained the spectral tilt value of the curvature power spectrum and the tensor-to-scalar ratio at $n_s-1=-0.032 \pm 0.006$ at $1\sigma$ and $r<0.11$ at $2\sigma$ respectively \cite{Ade:2015xua, Ade:2015lrj}.
The current resolution of the temperature and polarizartion anisotropies of the CMB probes, although unprecedented, 
has not been powerful enough to support or exclude the different BSM physics schemes. There are promising prospects that the proposed next generation CMB experiments, such as the LiteBIRD \cite{Matsumura:2016sri}, Core+ \cite{Finelli:2016cyd}, CMB-S4 \cite{Abazajian:2016yjj}, PRISM \cite{Andre:2013nfa}, PIXIE \cite{Kogut:2011xw},  will improve significantly on this direction. The sensitivity forecasts for $n_s$ and $r$ is of the order of $10^{-3}$ and such a measurement will account for a substantial leap forward at the observational side.

We aim at this work to show how one can systematically extract non-trivial information about the BSM physics via the CMB precision measurements.
We mostly focus on the supersymmetry since we consider it as a compelling 
BSM theory that remains elusive from the terrestrial colliders. A precise knowledge of the ($n_s, r$) values can indicate us
the duration of non-thermal phase after inflation and in this paper we use this information to examine whether different supersymmetry breaking schemes can fit in this picture of the early cosmic evolution.

From the 
experimental side, there is no signal that supports the supersymmetry hypothesis until today, see e.g. a recent analysis of searches at the LHC \cite{Aad:2016eki, Khachatryan:2016xvy}. The absence of signals arouses increasing concern that supersymmetry does not fully solve the hierarchy problem suggesting that supersymmetry, if realized, may lay at energy scales much higher than the TeV scale. Multi TeV supersymmetry implies that the Large Hadron Collider (LHC) at CERN may find no BSM signal and 
the fiducial BSM physics scenarios will remain elusive for an unspecified long time. 
However from the telescopic observational side, the increasing sensitivity of the CMB probes has opened up a rich phenomenological window to the ultra high energy scales of cosmic inflation and indirectly to the dark pre-BBN period.

Definitely, the idea that the CMB studies may probe energy scales well above the TeV is not a new one. There are numerous of seminal works in the literature that examine the impact of BSM physics, and in particular supersymmetry, on the CMB power spectrum mainly either  from the inflationary model building or from the dark matter perspective.   
However, successful inflation models can be consistently embedded into a supergravity framework often without any change in the inflationary dynamics since 
the inflationary trajectory may remain intact by the presence of additional supersymmetric fields that are efficiently stabilized. 
Moreover, it is often the case that studies of supersymmetric dark matter cosmology focus on the dark matter density parameter fitting, $\Omega_\text{DM}h^2=0.12$,  neglecting other features of the scalar power spectrum.

The degeneracy between supersymmetric inflation models and with their non-supersymmetric versions in terms of the $n_s(N)$ and $r(N)$ observables can break due to the different post-inflationary evolution.  The thermal evolution of a supersymmetric plasma is in general much different when supersymmetry is realized in nature \cite{Jungman:1995df}.
Actually, the null LHC results push the sparticles mass bounds to larger values that spoil the nice predictions of the thermal dark matter scenario \cite{Baer:2011ab}. 
Therefore, assuming that the LSP is part of the dark matter in the universe 
{\it the $\Omega_\text{LSP} h^2\lesssim 0.12$ constraint reconciles only 
with particular  
 radiation domination histories} which may greatly differ to the simple scenario of a single and smooth radiation phase after the inflaton decay.
An interesting point, that stimulates this work, is that the features of the radiation dominated phase 
depend on the details of the supersymmetry breaking patterns.

In order to extract information about the BSM supersymmetric scenarios from the $(n_s,r)$ precision measurements we utilize existing results on supersymmetric cosmology aiming at an analysis based on assumptions as minimal as possible.  We consider that the MSSM plus the gravitino is the necessary minimal set-up that gives the most conservative results.
We a priori consider the $T_\text{rh}$ and the supersymmetry breaking scale 
as unknown quantities.
We estimate the neutralino and gravitino LSP abundances by scanning the sparticle mass parameter space. As a rule of thumb we adopt the classification of quasi-natural, split and high scale supersymmetry when we scan the possible energy scales of supersymmetry breaking.
As expected, see e.g. \cite{Baer:1995nc, Baer:2014eja, Badziak:2015dyc}, we find that most of parameter space of supersymmetric theories yields an excessive dark matter abundance. 
Our perspective in this work is that the parameter space that yields an excessive dark matter abundance should not be faced as a cosmologically forbidden one but, on the contrary, as a parameter space that favours a different cosmic history for the very early universe. Namely, excessive LSP abundance implies either a low reheating temperature after inflation or low entropy production. Both cases have a non-trivial impact on ($n_s, r$) observables, see e.g. \cite{Easther:2013nga} for a relevant analysis on non-thermal neutralino dark matter and \cite{Maity:2018dgy} for a recent analysis on FIMP dark matter.

Departing from the minimal field content analysis, i.e. the MSSM, the overabundance problem in general deteriorates. 
Indeed, the dark matter abundance receives contributions from the perturbative and non-perturbative decay processes of the inflaton field \cite{Kofman:1997yn} and from thermal scatterings, thermal and non-thermal decays of fields coming from the supersymmetry breaking sector such as the messengers.
Extra fields can however {\itshape decrease} the DM abundance if they decay late and dominate the energy density of the early universe e.g. due to coherently oscillating scalars or scalars that cause thermal inflation. Such fields are rather common and well motivated in many BSM schemes such as supersymmetry; common examples are the moduli, supersymmetry breaking fields, the saxion, etc. 
Here we collectively label $X$ any of this sort of scalars and explicitly refer to it as {\itshape diluter}, since  what we actually measure on the CMB is the diluter impact on the expansion history. In our analysis, the {\it diluter} is the only field beyond the MSSM and gravitino that we consider. Finally, in order to perform a complete calculation of the spectral index value we consider the Starobinsky $R^2$ inflation model and we compare the $R^2$ inflation and $R^2$ supergravity inflation predictions by taking into account the effects of the post-inflationary phase.

Apparently one {\itshape cannot} exclude or verify supersymmetry by $n_s$ and $r$ precision measurement, nevertheless  one can indeed support the presence of BSM physics or, to put it differently, rule out the so-called {\itshape BSM-desert} hypothesis for a particular inflation model.
This is a minimal but undoubtedly an exciting possibility given the fact that terrestrial colliders probe only a small part of the vast energy scales up to the Planck Mass, $M_\text{Pl}$, and supersymmetry or any other BSM scale may lay anywhere in between. It is also exciting  to note that the terrestrial experiments, such as colliders and direct detection experiments,  are sensitive to low scale supersymmetry whereas the CMB observables are more sensitive to high scale supersymmetry. {\itshape Hence precision cosmology can offer us complementary constraints to the parameter space of the supersymmetric theories}. 
This prospect, though very challenging, is actually a feasible possibility.

The organization of the paper is the following. In section 2 we parametrize the uncertainty in the $n_s$ and $r$ values coming from  the unknown value of $N_*$ due to the dark pre-BBN era. We compute the shift in the spectral index and tensor-to-scalar ratio with respect to the dilution magnitude in a general BSM context.
In section 3 we overview key results of neutralino, gravitino and briefly the axino cosmology regarding the LSP yield, that are necessary for the estimation of the dilution magnitude. In section 4 we analyze the implications of various supersymmetry breaking patterns to the early universe cosmology and examine the features of the possible alternative cosmic histories. In section 5 the Starobinsky $R^2$ inflation is used as a specific example to demonstrate a full computation of the spectral index and tensor-to-scalar ratio shift. A comparison between the theoretical predictions of the $R^2$ and supergravity $R^2$ inflation is also performed.
In the last section we outline the main idea and the method proposed in this work and we comment on the future theoretical and observational prospects.


\section{CMB observables and the post-inflationary evolution}

It is convenient to expand the power spectra of the dimensionless curvature perturbation as
\begin{equation}
{\cal P}_{\cal R}(k)=A_s \left(\frac{k}{k_*}\right)^{n_s-1+(1/2)(dn_s/d\ln k)\ln(k/k_*)+(1/6)(d^2n_s/d\ln k^2)(\ln(k/k_*))^2+...}
\end{equation}
where $A_s$ is the scalar amplitude and the powers of the expansion are the scalar spectral index $n_s$, the running and the running of the $n_s$.
In general one can assume that the scale dependence of the spectral index to be given at leading order by the expression 
\begin{equation} \label{nsM}
n_s(k_*) =1 -\frac{\alpha}{N_*}\,,
\end{equation}
where $N_*$ is the number of e-folds remaining till the end of inflation after the moment the
pivot scale $k_*$ exits the Hubble radius, $N_*\equiv \int^{t_\text{end}}_{t_*} H dt=\ln(a_\text{end}/a_*)$.
The $N_*$ is a critical quantity that determines the $n_s$ value.
It carries the information of how much the observable $k^{-1}_*$ CMB scale 
has been stretched since the inflationary era. The uncertainty on the $N_*$ comes mainly from the post-accelaration stage and induces an uncertainty on the spectral index value given by the $n_s$ running  that for the Eq. (\ref{nsM}) reads
\begin{equation} \label{n_appr}
\Delta n_s = \alpha \frac{\Delta N}{N^2} = \frac{(1-n_s)^2}{\alpha} \Delta N\,.
\end{equation}
For $ \Delta N \sim 1-10$ the $\Delta n_s$ is of size ${\cal O}(1-10) \permil$ ,  that is within the accuracy of the future observations.

To explicitly estimate the $N_*$ value one relates the size of the scale $k^{-1}_*=(a_*H_*)^{-1}$, which exited the Hubble radius $H^{-1}_*$ during inflation, to the size of the present Hubble radius $H^{-1}_0$ \cite{Liddle:2003as},
\begin{equation} \label{k1}
\frac{k_*}{a_{0} H_{0}} = \frac{a_*}{a_\text{end}} \frac{a_\text{end}}{a_\text{BBN}} \frac{a_\text{BBN}}{a_\text{eq}} \frac{a_\text{eq}}{a_0}\frac{H_*}{H_\text{eq}}\frac{H_\text{eq}}{H_0}\,,
\end{equation}
where the subscripts refer to the time of horizon crossing ($*$), the time inflation ends (end), the time BBN takes place (BBN), the radiation-matter equality (eq) and the present time (0). 
We define $\tilde{N}_\text{dark}$ the  number of e-folds from the end of inflation until the beginning of the BBN
\begin{equation}
\tilde{N}_\text{dark} \equiv \ln \left(\frac{a_\text{BBN}}{a_\text{end}} \right) \equiv \frac{1}{3(1+\bar{w}_\text{dark})} \ln \frac{\rho_\text{end}}{\rho_\text{BBN}}\,,
\end{equation}
where $\bar{w}_\text{dark}$ stands for the average value of the equation of state parameter during the dark pre-BBN period, and $\bar{w}_\text{dark} \neq -1$ has been assumed.
We call this period {\itshape dark} due to the lack of observational evidences of the transition to the radiation dominated phase from the super-cooled conditions during inflation.
Unless exotic forms of matter are assumed, such as thermal inflation or stiff fluid domination,  we can estimate the maximum value of the $\tilde{N}_\text{dark}$ to be around $56$ for $\bar{w}_\text{dark}=0$ and the minimum to be around 41 for $\bar{w}_\text{dark}=1/3$. The observational uncertainty for temperatures $T\gtrsim 1$ MeV $\sim T_\text{BBN}$ \cite{Kawasaki:1999na} implies an uncertainty at the e-folds of inflation about $\Delta N \sim 15$. 
We can split the $\tilde{N}_\text{dark}$ into 
\begin{equation}
\tilde{N}_\text{dark}=\tilde{N}^\text{}_\text{rh} + \tilde{N}^\text{}_X + \tilde{N}^\text{}_\text{rad} 
\end{equation}
where $\tilde{N}^\text{}_\text{rh}=\ln(a_\text{rh}/a_\text{end})$ stands for the e-folds number of the postinflationary reheating period until the complete decay of the inflaton, $\tilde{N}^\text{}_\text{rad}$ 
the e-folds number of the radiation dominated era that preceded the BBN and $\tilde{N}^\text{}_X$ 
stands for the e-folds number that take place during the domination of an arbitrary $X$ field in the period after the decay of the inflaton and before BBN.

After plugging in the value for the ratio $a_\text{eq}H_\text{eq}/(a_0H_0)$, the relation (\ref{k1}) is recast into \cite{Ade:2015lrj} 
\begin{equation}
N_*\approx 66.7 -\ln\left( \frac{k_*}{a_0 H_0}\right) +\frac14 \ln\left( \frac{V^2_*}{M^4_\text{Pl} \rho_{\text{end}}}\right) -\frac{1-3\bar{w}_\text{dark}}{4} \tilde{N}_\text{dark}\,. 
\end{equation}
Utilizing the relation ${\cal P}_{\cal R}(k_*)={V_*}/{(24\pi^2 \epsilon_* M^4_\text{Pl})}=A_s$ and after substituting numbers for the the ratio $k_*/(a_0H_0)$ we get
\begin{equation} \label{N1}
N_*\approx 60.8 + \frac14 \ln\epsilon_* +\frac14 \ln\frac{V_*}{\rho_\text{end}} - \Delta N_\text{dark}\,. 
\end{equation}
We adopted the {\itshape Planck} collaboration pivot scale, $k_*=0.002 \text{Mpc}^{-1}$ and the measured value $\ln(10^{10}A_s)=3.089$  \cite{Ade:2015xua, Ade:2015lrj}.
We also introduced the $\Delta N_\text{dark}$ factor to mark explicitly the uncertainty of the dark pre-BBN era on the $N_*$ value, 
\begin{equation} 
\Delta N_\text{dark} \equiv  \frac{1-3\bar{w}_\text{dark}}{4} \tilde{N}_\text{dark} = \Delta N_\text{rh}+ \Delta N_X + \Delta N_\text{rad}\,.
\end{equation}
We have split  the $\Delta N_\text{dark}$ into the contributions from the inflationary reheating, the X-domination and the pre-BBN radiation domination period.
It is $\Delta N_\text{rad}=0$ since $\bar{w}_\text{rad}=1/3$ and
\begin{equation} \label{DNb}
\Delta N_\text{rh}=\frac{1-3\bar{w}_\text{rh}}{12(1+\bar{w}_\text{rh})}\ln\left( \frac{\rho_\text{end}}{\rho_\text{rh}}\right)\,,\quad\quad\quad
\Delta N_\text{X}=\frac{1-3\bar{w}_\text{X}}{12(1+\bar{w}_\text{X})}\ln\left( \frac{\rho^\text{dom}_X}{\rho_X^\text{dec}}\right)\,,
\end{equation}
where $\rho_X^\text{dec}$ is the energy density of the thermal plasma right after the decay of the scalar X. In principle, for a concrete and predictable inflationary model the $\bar{w}^\text{}_\text{rh}$ and the reheating temperature after inflation can be estimated and hence the $\Delta N_\text{rh}$. The crucial quantity is the decay rate $\Gamma_\text{inf}$ of the inflaton which determines the reheating temperature. Assuming that the decay and the thermalization occur instantaneously  at the time $\Gamma^{-1}_\text{inf}$ then the reheating temperature is found by equating (and omitting order one coefficients) $\Gamma_\text{inf}=H$,
\begin{equation}
T^\text{}_\text{rh} =\left(\frac{\pi^2}{90}g_{*\text{rh}}\right)^{-1/4} \sqrt{\Gamma_\text{inf} M_\text{Pl}}\,.
\end{equation}
The maximum temperature possible is achieved in the instant reheating scenario. Apparently when $T_\text{rh}=T_\text{max}=\rho^{1/4}_\text{end}(30/\pi^2g_{*\text{rh}})^{1/4}$ it is $\Delta N_\text{rh}=0$. Note that the $N_*$ has a logarithmic dependence on $g_{*\text{rh}}$, with $g_{*\text{rh}}$ being the effective number of relativistic species upon thermalization.

It is however well possible that after the inflaton decay the evolution of the universe could have been episodic with additional reheating events after inflation. Hence the cosmic thermal era
could have started after the last reheating stage before primordial nucleosynthesis caused by other than the inflaton scalar field, for instance a modulus or a flaton \cite{Lyth:1995ka} that we collectively label $X$.  
Here, we prefer to remain agnostic about the identity of $X$ but we do utilize its property to cause efficient dilution and low entropy production. 
The $X$ can dominate the energy density of the universe over radiation due to the slower redshift of its energy density stored. It is  $\rho_X \propto a^{-3}$ for a scalar condensate that coherently oscillates in a quadratic potential and $\rho_X \approx \text{constant}$ for a scalar field with sufficiently flat potential that causes thermal inflation.


\subsection{The shift in the scalar spectral index and tensor-to-scalar ratio due to late entropy production}

Let us now estimate the impact of the $X$ domination era on the spectral index value.
We call $N^{(\text{th})}$ and  $n^{(\text{th})}_s$ the {\it thermal reference  values}, that is the e-folds number and the spectral index values respectively if there is no late entropy production after the inflaton decay, i.e. dilution effects. It is at leading order
\begin{equation} \label{N*}
N_*=N^{(\text{th})}-\Delta N_X \,,\quad \quad n^{(\text{th})}_s=1-\alpha/N^{(\text{th})} \,,
\end{equation}
where, following Eq. (\ref{N1}),
\begin{equation} \label{Nth}
N^{(\text{th})}= 60.8 +\frac14 \ln \epsilon^{(\text{th})} +\frac14 \ln \frac {V^{(\text{th})}}{\rho_\text{end}} - \Delta N^\text{}_\text{rh}\, .
\end{equation} 

At leading order the scalar tilt is generally given by the  equation (\ref{nsM}). 
Since precision is expected to increase in the future it is worthwhile to consider next-to-leading corrections. 
Due to the large number of inflationary models \cite{Martin:2013tda} there is no common form for the next-to-leading term \cite{Martin:2016iqo}. A phenomenological way to parametrize it is based on the large $N$ expansion 
\begin{equation} \label{nsM_next}
n_s =1 -\frac{\alpha}{N}+\frac{\beta(N)}{N^2}  +{\cal O}\left(\frac{1}{N^3}\right)    
\,.
\end{equation}
The parameters $\alpha$ and $\beta(N)$ are determined only after a particular inflation model is considered. In principle the parameter $\alpha$ can also be a slowly varying function of $N$ \cite{Creminelli:2014nqa}.  In addition the expansion (\ref{nsM_next}), for some inflation models, 
may involve parameters of the potential \cite{Martin:2016iqo}.
Here we assume that $\alpha$ is a constant and absorb possible complicated behaviors in the arbitrary $\beta(N)$ function.
In section 5 we will explicitly estimate the shift
in the spectral index for the Starobinsky $R^2$ inflation model where the parameters $\alpha$  and $\beta(N)$ have a  particular form.

If $\Delta N_X \neq 0$, after Taylor expanding the $n_s\left(N^{(\text{th})}-\Delta N_X\right)$, the spectral index $n_s^{(\text{th})}=n_s(N^{(\text{th})})$ value  is shifted by an amount $\Delta n_s\equiv n_s-n_s^\text{(th)}$,
\begin{equation} \label{Dn_F}
\Delta n_s =  - \left(1-n_s^{(\text{th})}\right) \left[ \frac{\Delta N_X}{N^{(\text{th})}} + \left( \frac{\Delta N_X}{N^{(\text{th})}}\right)^2 
+\left(\frac{\Delta N_X}{N^{(\text{th})}}\right)^3    \right] 
 + F_\beta\left(\Delta N_X, N^{(\text{th})}\right)
\end{equation}
where $1-n_s^{(\text{th})}= \alpha/N^{}-\beta(N^{})/N^2|_{N=N^{(\text{th})}}$ and 
\begin{equation} \label{Fb}
\begin{split}
 F_\beta\left(\Delta N_X, N^{(\text{th})}\right)= & \left(\beta - \beta'N^{}\right) \frac{\Delta N_X}{N^3}\, +\,2\left(\beta - \beta' N^{} +\frac14 \beta'' N^2\right) \frac{\Delta N_X^2}{N^4}\, + \\
& \left. 3\left(\beta - \beta' N + \frac13 \beta'' N^2 - \frac{1}{18} \beta''' N^3 \right) \frac{\Delta N_X^3}{N^5} \,\right|_{N=N^{(\text{th})}}\,.
\end{split}
\end{equation}
The $"'"$ denotes $d/dN$ and $\beta, \beta'$, $\beta''$, $\beta'''$ are estimated at $N=N^{(\text{th})}$. 
In the above expressions, given than $\Delta N_X>1$ and $\Delta N_X/N^{(\text{th})}<1$, terms of order ${\cal O}\left({\Delta N_X^4}/N^6\right)$ 
and smaller have been neglected. 
We have also assumed that the terms in the parentheses in Eq. (\ref{Fb}) are roughly of order $\beta$. Otherwise, if $\beta', \beta'', \beta''' \gg 1$, 
the $F_\beta$ correction can be important, however such a behavior is not found in any of the known universality classes \cite{Garcia-Bellido:2014gna}. 
One can see that the next-to-leading correction $\beta(N)/N^2$ is at most of $\permil$ accuracy and for $\alpha \Delta N_X > \beta$ the contribution to the spectral index shift is found to be subdominant with respect to the $\alpha$-dependent terms.

In order to specify the $\Delta N_X$, elements of the $X$ scalar cosmic evolution have to be specified.
When the scalar $X$ {\itshape coherently oscillates} about the minimum of a effectively quadratic potential  it is $\bar{w}_X=0$. In such a case, at the cosmic time $t_X^\text{dom} \ll \Gamma^{-1}_X$ the energy density of $X$ is larger than that of the plasma and the universe enters a scalar dominated era that dilutes any pre-existing abundances of the relativistic degrees of freedom at the time of the $X$ decay. The $X$ field decays and reheats the universe with temperature $T_X^{\text{rh}}\equiv T_X^{\text{dec}}$. Considering instant decay of the scalar $X$, the dilution magnitude is estimated to be
\begin{equation} \label{DX}
D_X  \equiv 1+\frac{S_\text{after}}{S_\text{before}} =1+ \frac{g_s(T_X^{\text{dec}})}{g_*(T_X^{\text{dec}})} \frac{g_*(T_X^\text{dom})}{g_s(T_X^\text{dom})}\frac{T_X^\text{dom}}{T_X^{\text{dec}}}\,  \simeq \, \frac{T_X^\text{dom}} {T_X^\text{dec}} \geq 1
\end{equation}
where $S_\text{before}$ and $S_\text{after}$ denote the entropy density right before and after the decay of the X field. The $g_*$ and $g_s$ count the total number of the effectively massless degrees of freedom for the energy density and entropy respectively and can be taken to be  approximately equal.
The $T_X^\text{dec}$ is the temperature that the $X$ scalar reheats the universe at the time $H^{-1}\simeq \Gamma^{-1}_X$. 
It is $D_X =1$ when no dilution takes place.
Overall, the size of the $\Delta N_X$ due to the $X$ scalar domination reads 
\begin{equation} 
\Delta N_X=\frac14 \tilde{N}^\text{}_X = \frac{1}{12} \ln \frac{\rho^\text{dom}_X}{\rho^\text{dec}_X}
\end{equation}
where we considered that $\bar{w}_X=0$. After plugging in the dilution magnitude we get
\begin{equation} \label{DN1}
\Delta N_X=\frac{1}{3} 
\ln \left[\left(\frac{g_*(T^\text{dom}_X)}{g_*(T^\text{dec}_X)}\right)^{1/4} D_X \right] \equiv \frac13 \ln\tilde{D}_X \,. 
\end{equation}
The maximum value of the $\Delta N_X\sim 15$ is achieved when $\tilde{N}^\text{}_\text{rh}\rightarrow 0$ and $\tilde{N}^\text{}_\text{rad}\rightarrow 0$. 
 This case corresponds to the maximum dilution scenario where the $X$ field oscillations dominate the energy density of the universe right after the end of high scale inflation until the onset of BBN. 
The $\Delta N_X=0$  case corresponds to an uninterrupted radiation phase following the post-inflationary reheating. 
If someone assumes the presence of $X$ matter with exotic barotropic parameter the $\Delta N_X$ limit values can be extended. 

Substituting $\Delta N_X= \frac13 \ln \tilde{D}_X$ 
in the expansion (\ref{Dn_F}) 
we obtain the shift in the spectral index, with accuracy $|\Delta n_s|/n_s \lesssim \, 1 \permil$,  due to a post-inflationary dilution of the thermal plasma 
\begin{equation} \label{Dn}
\Delta n_s = - \left(1-n^{(\text{th})}_s \right)^2 \,\frac{\gamma^{}}{3 \alpha} \ln \tilde{D}_X 
\left [ \, \sum_{p=0}^2 \left( \gamma^{} \frac{1-n^{(\text{th})}_s}{3 \alpha} \ln \tilde{D}_X \right)^p - \frac{\beta \gamma^2}{\alpha^2}\left(1-n_s^{(\text{th})}\right) +\frac{\beta'\gamma}{\alpha} \right]\,,
\end{equation}
where  $1-n_s^{(\text{th})}= \alpha/N^{}-\beta(N^{})/N^2|_{N=N^{(\text{th})}}$, $\beta=\beta(N)|_{N=N^{(\text{th})}}$, $\beta'=\beta'(N)|_{N=N^{(\text{th})}}$  and $\gamma^{}=1+\beta(N)[(1-n_s)N^2]^{-1}|_{N=N^{(\text{th})}}$. 
Notice that at leading order the (\ref{Dn}) reads $\Delta n_s=- \alpha_s \,\Delta N_X$, 
where $\alpha_s=(1-n^{(\text{th})}_s)^2 \gamma/\alpha$ is the running of the spectral index at $N^{(\text{th})}$. We also mention that the three last terms in the brackets of the above equation can be neglected without significant cost in the $\permil$ accuracy. 

Plugging in the thermal reference value $n_s^{(\text{th})}$  that a given inflation model yields, the expression (\ref{Dn}) returns the shift in the spectral index due to a pre-BBN dilution of the thermal plasma. 
We see that the $\Delta n_s$ is {\itshape negative} which means that the spectrum tilt becomes more red when dilution of the radiation plasma takes place; this behavior is illustrated in Fig. 1. 
The precision of the expression (\ref{Dn})  is sufficiently good, i.e one per mile, even for the extreme case $\Delta N_X \sim 15$ or $D_X\sim 10^{20}$.

Apart from scalar condensates, several BSM construction, mostly supersymmetric ones, predict the presence of singlets under the Standard Model symmetries that have a relatively flat potential. Such fields can realize the {\itshape thermal inflation} scenario and are generally called flatons \cite{Lyth:1995ka}. Due to Yukawa interactions the flaton can be trapped at the origin of the field space by thermal effects. At some temperature that we denote $T^\text{dom}_X$ the vacuum energy $V_0$ of the flaton dominates over the background radiation energy density and a period of thermal inflation starts. Thermal inflation ends at the temperature $T_2$ when the thermal trap has become too weak and the flaton field starts oscillating about its zero temperature minimum. The flaton finally decays at the temperature that we denote $T^{\text{dec}}_X$ and we consider instant reheating. 
The dilution magnitude due to the {\it flaton $X$ domination} is 
\begin{equation} \label{FD}
  D_X^\text{FD} \, \simeq \,  1+\frac{(T^\text{dom}_X)^4}{T_2^3 T_X^\text{dec}}\, \simeq\, \frac{(T^\text{dom}_X)^4}{T_2^3 T_X^\text{dec}}  \,.
\end{equation}
Respectively the $\Delta N_X$ value for flaton domination is 
\begin{equation} \label{DN2}
\left. {\Delta N_X} \right|_\text{FD}\, = \,  \ln\left[\frac{g^{1/4}_*(T^\text{dom}_X)}{g^{1/4}_*(T_2)}\frac{T^\text{dom}_X}{T_2} \right] + \frac13 \ln \left[\frac{g^{1/4}_*(T^\text{dom}_X)}{g^{1/4}_*(T^\text{dec}_X)}\frac{T^\text{dom}_X}{T_X^\text{dec}} \right] \equiv \frac13 \ln\tilde{D}^{\text{FD}}_X \,,
\end{equation} 
and each term can be written in a compact form $\left. \left.\left. {\Delta N_X} \right|_\text{FD}\, = \, {\Delta N_X} \right|_\text{TI} + {\Delta N_X} \right|_\text{SC}$, where TI and SC stand for thermal inflation and scalar condensate respectively.

The ratio of the relativistic degrees of freedom accounts for a small correction and one can see that it is actually $\left. \Delta N_X\right|_{\text{FD}} \simeq  \ln D^{\text{FD}}_X/3$. 
The dilution magnitude maximizes when 
 the thermal inflation phase is followed by a scalar condensate domination phase, e.g. when the flaton field decays very slowly. 
The $\Delta N$ due to thermal inflation  has an upper bound $\tilde{N}_\text{TI}\sim 10$ 
in order that the cosmological density perturbation remain intact. Nevertheless, the dilution magnitude can be many orders of magnitude larger than the dilution caused by a scalar condensate domination (\ref{DX}) and it can efficiently dilute any overabundant relic such as dark matter particles.
Essentially, the shift in the spectral index due to thermal inflation, $\left. \Delta n_s \right|_\text{FD}$, is given again by the Eq. (\ref{Dn}) simply by replacing the $\ln\tilde{D}_X$ with the $\ln\tilde{D}^{\text{FD}}_X$, see Fig 1. It is remarkable that the shift in $n_s$ due to period of thermal inflation can resurrect ruled out inflationary models such as the minimal hybrid inflation in supergravity \cite{Dimopoulos:2016tzn}. 
 
Finally, let us comment on the shift in the tensor-to-scalar ratio. The phenomenological parametrization of the scalar tilt $n_s=1-\alpha/N$ implies that the first slow roll parameter $\epsilon=-\dot{H}/H^2$ writes \cite{Creminelli:2014nqa}
\begin{equation} \label{eN}
\epsilon(N)=\frac{1}{2(\alpha-1)^{-1}N+AN^\alpha}\,,
\end{equation}
where $A$ an integration constant coming from the differential equation $\epsilon+d\ln\epsilon/dN=\alpha/N$. 
At first order in slow roll we have $r=16\epsilon$ and the shift in the tensor-to-scalar ratio due to a non-thermal phase is $\Delta r=r(N^\text{(th)}-\Delta N_X)-r(N^\text{(th)}) \simeq -r'(N^\text{(th)}) \Delta N_X$, i.e.
\begin{equation} \label{dr}
\Delta r= \frac{\left(r^\text{(th)}\right)^2}{16}\left[2(\alpha-1)^{-1} +\alpha A N^{\alpha-1} \right]\, \Delta N_X\, ,
\end{equation}
where $r^\text{(th)}=r(N^\text{(th)})$. 
For $\Delta N_X=\ln \tilde{D}_X/3$, either due to a scalar condensate domination or thermal inflation, the relation $\Delta r=\Delta r(\tilde{D}_X)$ is obtained.
The scaling (\ref{eN}) depends on the potential that implements inflation.
Different potentials yield different values for $\alpha$ and $A$. Moreover, accuracy of order $\Delta r\sim 10^{-4}$ requires to go beyond the approximate relation $r=16\epsilon$ and consider corrections at second order in slow roll.  
In section 5 we will explicitly estimate the $\Delta r$ for the Starobinsky $R^2$ inflation model with the next-to-leading order corrections taken into account.
The general conclusion is that, according to Eq. (\ref{dr}), a non-thermal phase with $\bar{w}_X<1/3$ and duration $\tilde{N}_X=[(1-3\bar{w}_X)/4]^{-1} \Delta N_X$  {\it increases}  the tensor-to-scalar ratio value. 
\\
\\
Summarizing, the duration of a non-thermal phase is encoded in the number of e-folds $N$ between the moment a relevant mode exits the horizon and the end of inflation. 
\begin{figure}
\centering
\begin{tabular}{cc} 
{} \includegraphics [scale=.60, angle=0]{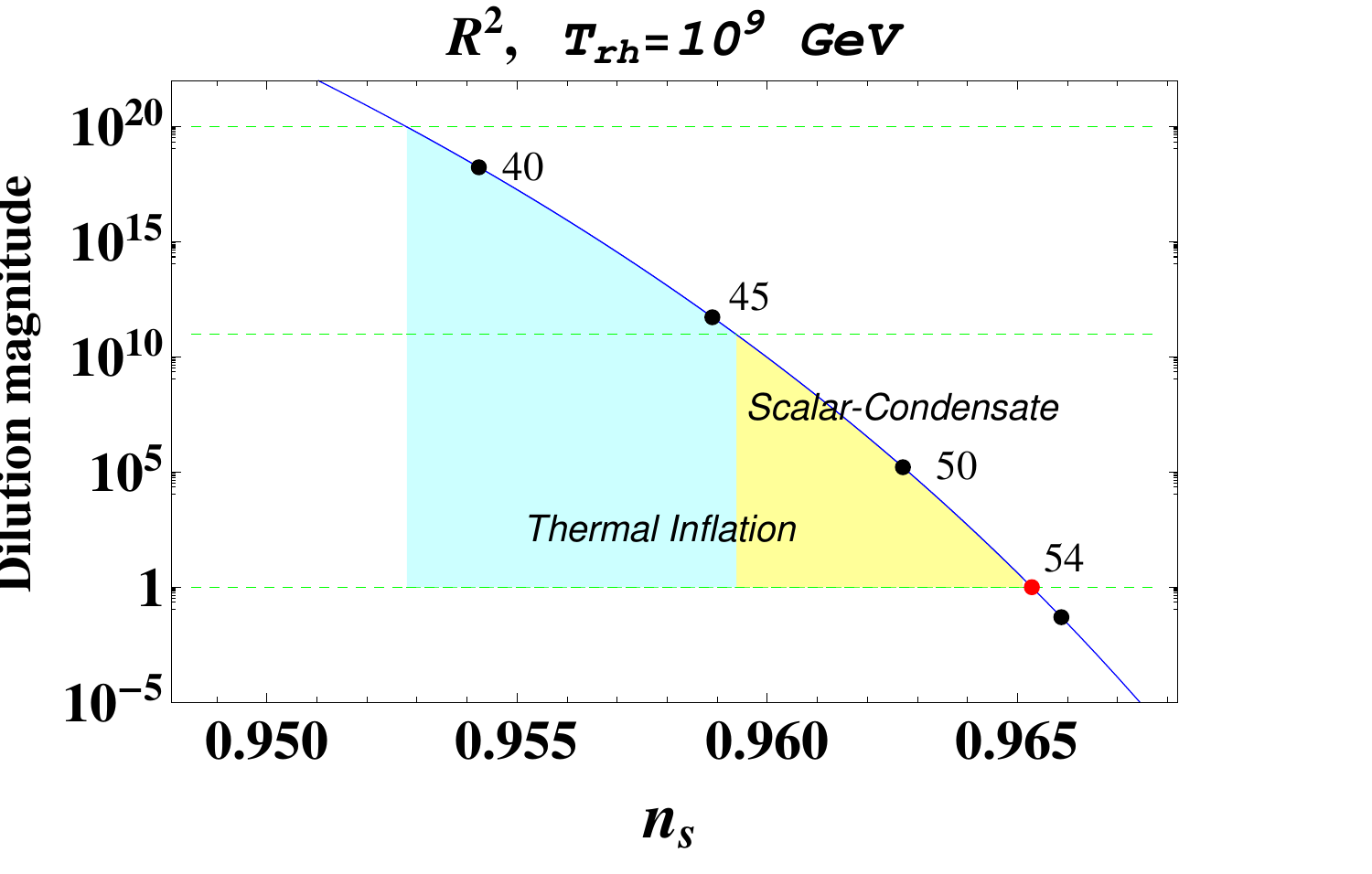} \,
{} \includegraphics [scale=.60, angle=0]{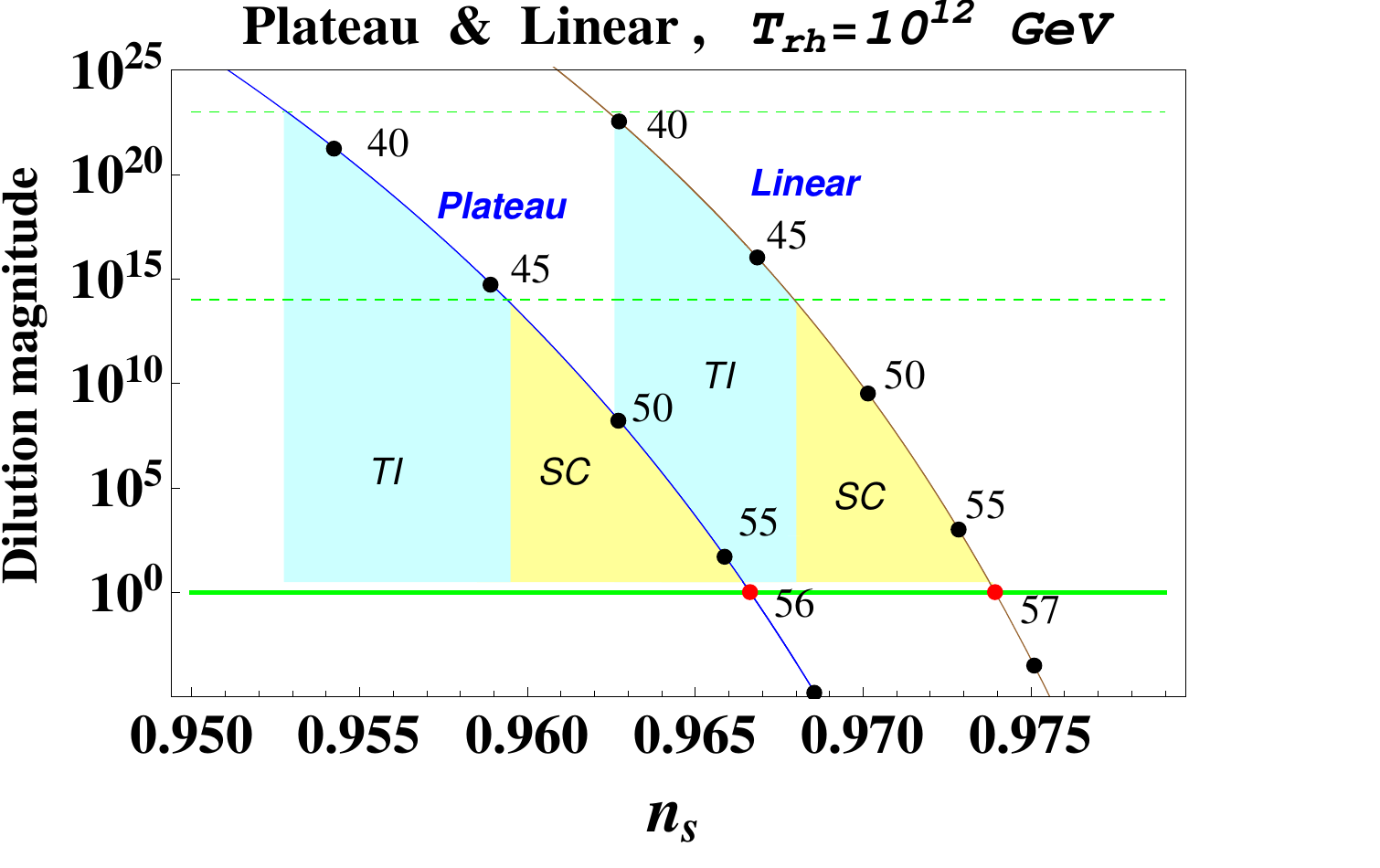}  \\
\end{tabular}
\caption {\small {The shift in the spectral index value and the dilution magnitude $D_X$ due to scalar condensate domination (SC) and due to thermal inflation (TI) for the Starobinsky $R^2$ inflation (left panel), general plateau and linear inflationary potentials (right panel). The maximum number of the dilution is given by the ratio $T_\text{rh}/T_\text{BBN}$ for scalar condensate domination and the $\left. \Delta N_X\right|_\text{TI} \lesssim 10$ constraint  for thermal inflation. The red dots show the e-folds number if there is no entropy production after infaton decay. It is $N^\text{(th)}\simeq 54$ for $R^2$ inflation and $N^\text{(th)}\simeq 56,\,57$ for the general plateau and linear potential respectively (red dots). Order ${\cal O}(1)$ corrections to the dilution magnitude are expected due to the uncertainty at the number of the relativistic degrees of freedom at ultra high energies.}}
\end{figure}
If the radiation domination era, where $w=1/3$, initiates at the moment of the complete inflaton decay and continues without break until the BBN epoch then the e-folds number, called here thermal e-folds number $N^\text{(th)}$, can be explicitly  determined by the dynamics and the full interactions of the inflaton field. 
If not, a non-thermal phase changes the aforementioned e-folds by the amount $\Delta N_X \sim \ln D_X/3$. A dilution of size $D_X=20$ translates into $\Delta N_X\sim 1$ and a prolonged dilution e.g. of size $D_X=10^{13}$ into  $\Delta N_X\sim 10$. 
In order to estimate the shift in the spectral index and the tensor-to-scalar ratio one has to know the $N^\text{(th)}$ that is given by the Eq. (\ref{Nth}). This is possible only after an inflationary model and the parameters describing reheating are chosen. Then from Eq. (\ref{nsM_next})  the $n^\text{(th)}_s=n_s(N^\text{(th)})$ and the $n_s=n_s(N^\text{(th)}-\Delta N_X)$ can be estimated and hence the spectral index shift $\Delta n_s$, given by the Eq. (\ref{Dn_F}) or (\ref{Dn}), is obtained. 
In Fig. 1 we illustrate the shift in the spectral index due to a non-thermal phase that is implemented after reheating and before BBN. In the left panel we considered the Starobinsky $R^2$ model that predicts $T_\text{rh}\sim 10^9$ GeV \cite{Takeda:2014qma}, and in the right panel a Starobinsky-like potential with non-gravitational interactions and a linear potential $V\propto \phi$ both characterized by a fiducial reheating temperature $T_\text{rh}=10^{12}$ GeV. The knowledge of these inflaton features enables the explicit calculation of the  $n_s^\text{(th)}$ value, that corresponds to the red dots in the plots.
A scalar condensate domination or thermal inflation shifts the spectral index value according to the formula (\ref{Dn}) as illustrated in the Fig. 1.

From a more bottom-up approach, the postulation of a non-thermal phase during the pre-BBN era is not enough to determine the $\Delta n_s$ and $\Delta r$. 
Although a rough estimation of the spectral index shift can be done by the approximate expression (\ref{n_appr})  
the result is far from accurate and cannot consistently constrain the early universe cosmic history.  The best method is to choose an inflation model that is in accordance with a particular BSM description of the early universe (e.g. a supersymmetric, stringy or modified gravity framework) and estimate the $\Delta n_s$ and $\Delta r$ according to the pre-BBN cosmology implied by the BSM theory at hand.  Examples of BSM cosmic processes connected with the expansion history of the universe are the dark matter production and the baryogenesis processes. 
In the following we will consider the supersymmetric BSM scenario and determine features of the pre-BBN cosmology, such as possible non-thermal stages, that allow the accommodation of different supersymmetry breaking schemes {\it assuming that the LSP is part of the dark matter in the universe}. We will estimate the minimum dilution size dictated by the requirement $\Omega_\text{LSP}h^2 \leq 0.12$ and determine the expected shift in the spectral index and tensor-to-scalar ratio when a particular inflation model, which in section 5 is the Starobinsky $R^2$ model, complements the description of the early universe evolution.


\section{Supersymmetric dark matter 
cosmology}

In the previous section we computed the shift in the spectral index and tensor-to-scalar ratio due to post-inflationary entropy production. 
The fact that the present universe contains dark matter with relic density $\Omega_\text{DM}h^2=0.12$  relates the amount of the dilution with the dark matter production. In this section, focusing on TeV and especially multi-TeV supersymmetric scenarios, we will overview the expected LSP yield. We will stress out that the dilution is generally required, hence the CMB inflationary observables should be non-trivially influenced by the post inflationary expansion history of the supersymmetric universe.

There are several fundamental theoretical reasons to believe that supersymmetry is a symmetry of nature. 
For the devotee of supersymmetry the central question is the scale that supersymmetry is realized.  
The direct superpartner LHC-limits for all colored sparticles exceed $1.5$ TeV and suggest that we should depart from scenarios with natural supersymmetry paying the price of pushing the amount of tuning at the MSSM to less than $0.5-1$ percent level. 
However, the absence of BSM signals in the LHC rules out {\itshape only} the electroweak scale supersymmetry and not supersymmetry in general. 

BSM physics scenarios with unnatural supersymmetry are still very appealing.  Gauge coupling unification, the presence of a stable dark matter particle, the possible baryogenesis processes and the stringy UV completion of the low energy theories do not link SUSY with the electroweak scale. Supersymmetry may appear at higher energy scales.
In Ref. \cite{Bagnaschi:2014rsa}, different supersymmetry breaking scenarios have been categorized according to the mass spectrum features into three representative cases: i)  {\it Quasi-natural supersymmetry}, in which supersymmetric particles are heavier than the weak scale, but not too far from it, about in the $1-30$ TeV range. ii) {\it Split supersymmetry},  in which only the scalar supersymmetric particles have masses of the order
of $\tilde{m}$, while gauginos and higgsinos are lighter, possibly with masses near the weak scale \cite{Wells:2003tf, ArkaniHamed:2004fb, ArkaniHamed:2004yi}. There are also the Mega-Split \cite{Benakli:2015ioa} or Mini-Split\cite{Arvanitaki:2012ps} scenarios. iii) Finally the {\it High-Scale supersymmetry}, see e.g \cite{Fox:2005yp, Hall:2009nd} in which all supersymmetric particles have masses around a common
scale $\tilde{m}$, unrelated to the weak scale.
The $\tilde{m}$ is constrained by the Higgs mass value according to the details, of each supersymmetry breaking scenario. Roughly in the Split supersymmetry the maximum value allowed for $\tilde{m}$ is $10^8$ GeV when $\tan\beta$ is small, while in the High Scale supersymmetry the $\tilde{m}$ value can be up to $10^{12}$ GeV.

For our analysis it is critical that the LSP is {\it stable}. The stability of the LSP dark matter is assured by the presence of a discrete symmetry of the supergravity Lagrangian, the $R$-parity. If the R-parity is violated then the cosmological constraint $\Omega_\text{DM} h^2=0.12$  is raised for the LSP.
Although $R$-parity violating models have been actually constructed and have interesting phenomenological implications \cite{Barbier:2004ez},  there are strong arguments based on GUT models that support the $R$-parity conservation even when the scale of supersymmetry breaking is well above the electroweak scale \cite{Acharya:2014vha}. These results motivate us to assume that the LSP lifetime is much larger than the age of the universe and thus the LSP is constituent  of the dark matter.

Given the supersymmetry breaking scheme the stability of the LSP puts strong constraints on the thermal history of the universe. 
In the following subsections we overview the basic relevant cosmological aspects and results of the gravitino and neutralino LSP scenarios necessary for the goals of our analysis.


\begin{figure} \label{f1}
\centering
\begin{tabular}{cc} 
{(a)} \includegraphics [scale=.72, angle=0]{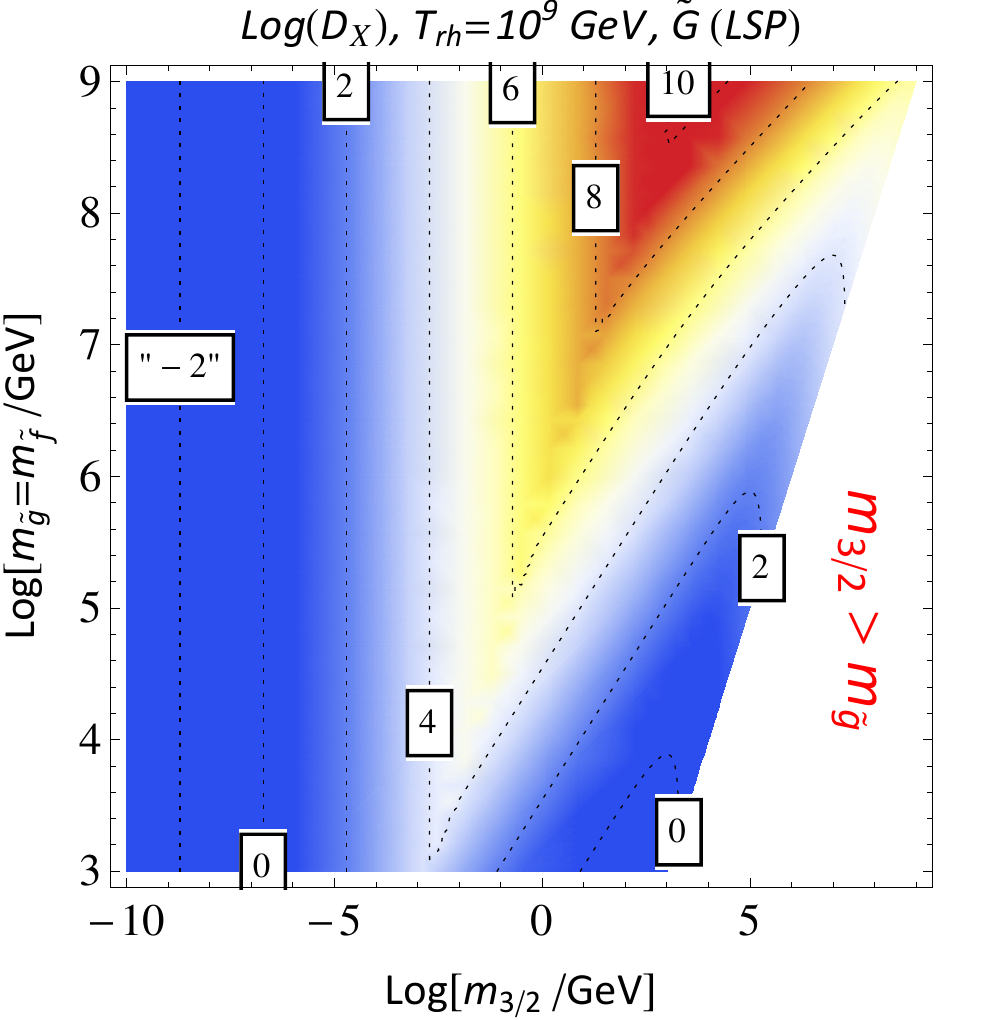} \quad
{(b)} \includegraphics [scale=.72, angle=0]{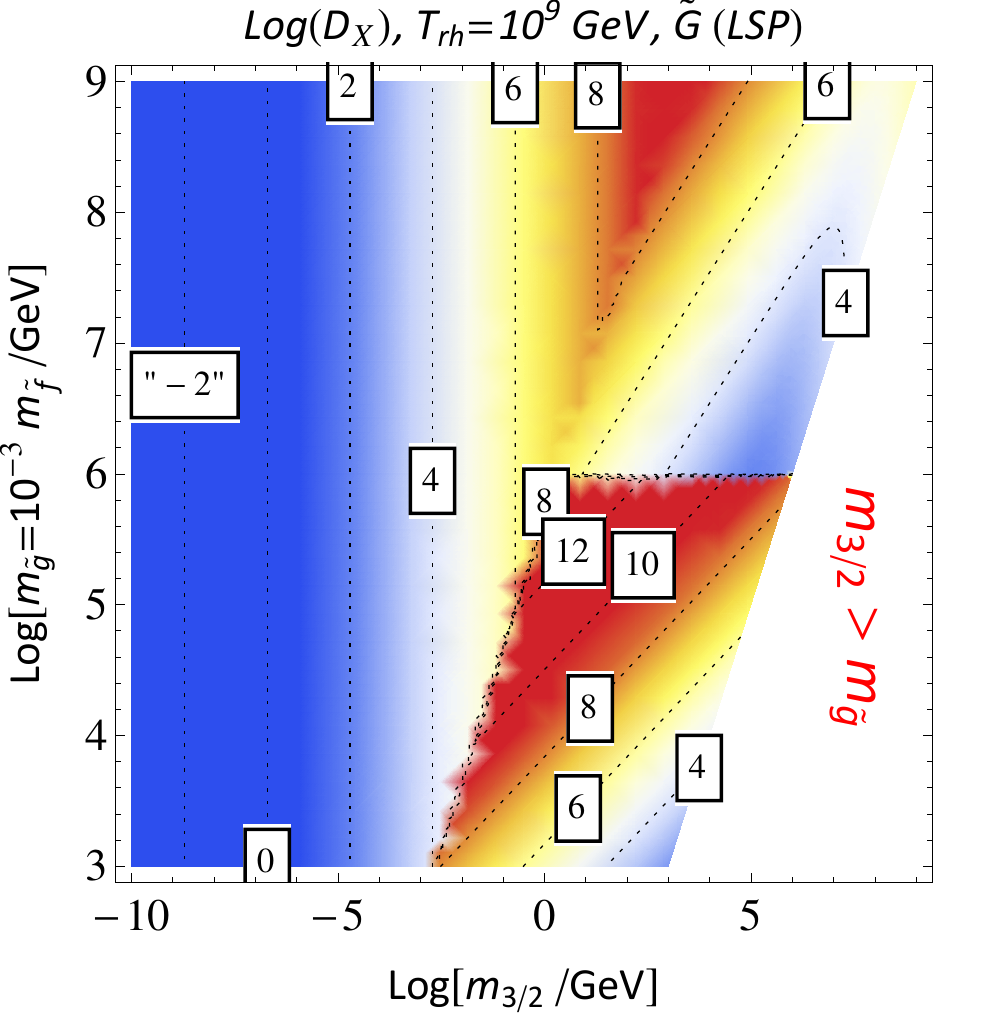}  \\
\end{tabular}
\caption{\small{ Density and contour plot of the decadic logarithm of the {\itshape required} dilution for $T_\text{rh} = 10^9$ GeV reheating temperature after inflation and gravitno the stable LSP. In the left panel degenerate spectrum for the sfermions and gauginos was considered, $m_{\tilde{f}}=m_{\tilde{g}}$, while in the right panel there is a split spectrum with $m_{\tilde{f}}=10^3 m_{\tilde{g}}$.  Thermal production of helicity $\pm 3/2$ and $\pm 1/2$ gravitinos from scatterings in the plasma, 
non-thermal production from decays of sfermions and the NLSP to helicity $\pm1/2$ gravitinos have been taken into account. The contributions to the gravitino abundance have been conditionally added, i.e. in the parts of the contour that thermal equilibrium is achieved the total abundance is replaced by the thermal one. 
The magnitude of the logarithm of the required dilution is given by the contour numbers onto the density plot. 
Negative numbers correspond to underabundance, hence to no dilution contour area. 
}}
\end{figure}

\subsection{Gravitino dark matter}

The gravitino is the supersymmetric partner of the graviton in supergravity and it can acquire a mass in the range of ${\cal O}(\text{eV}-\tilde{m})$.
The gravitino is naturally the LSP in gauge mediated supersymmetry breaking models (GMSB), see \cite{Giudice:1998bp} for a review, and possibly it is the LSP in Split and High scale supersymmetry frameworks. 
The relic density of the gravitinos $\Omega_{3/2} h^2$, which can be thermal or non-thermal, receives contributions from many sources.
\begin{description}
	\item [] {\itshape {Thermal gravitinos (freeze out)}}:  From scatterings (i) with the MSSM plasma, (ii) with the thermalized messenger fields.
	\item [] {\itshape {Non-thermal gravitinos (freeze in)}}: 	From (i) thermal scatterings in MSSM and messengers plasma, (ii)  decays of sfermions and the NLSP,  (iii) decays of the messenger fields, (iv) perturbative and non-perturbative decay of the inflaton field, (v) decay of moduli fields.
\end{description}

The less model independent estimation of the $\Omega_{3/2}$ is achieved when only the MSSM sector is considered. 
The gravitino number density $n_{3/2}$ in the thermalized early universe evolves according to the Boltzmann equation \cite{Kolb:1990vq}.
A key quantity is the 
 the gravitino production rate, $\gamma_\text{sc}$, in scatterings with thermalized Standard Model particles and sparticles 
\begin{equation} \label{Grsc}
\gamma_\text{sc}\sim 0.1 \frac{T^6}{M^2_\text{Pl}} \left(1+\frac{ m^2_{\tilde{g}}(\mu)}{3 m^2_{3/2}} \right)\equiv 0.1 \frac{T^6}{M^2_\text{Pl}} \hat{\gamma}_\text{sc} \,. 
\end{equation}
The gravitinos obtain a thermal distribution  via interactions with the MSSM for $T_\text{rh}>T^\text{f.o.}_{3/2}\sim 2\times 10^{14} \text{GeV} \left( {m_{3/2}}/{\text{GeV}} \right)^2 \left( {\text{TeV}}/{m_{\tilde{g}_3}} \right)^2$, where $m_{\tilde{g}_3}$ is the gluino mass evaluated at the reheating temperature, see Eq. (\ref{Oth1}).
If the reheating temperature is below the $T^\text{f.o.}_{3/2}$ the gravitino yield from MSSM thermal scatterings is $Y^\text{MSSM(sc)}_{3/2}\sim 10^{-3} \left(T_\text{rh}/T^\text{f.o.}_{3/2}\right)$.
Furthermore,  the heavier MSSM sparticles are unstable and will decay to gravitinos. The decay width into gravitinos is nearly the same for both gauginos and sfermions
\begin{equation} \label{ssmd}
\Gamma^\text{MSSM}(\tilde{i} \rightarrow i\,\tilde{G}  )\simeq\frac{1}{48\pi}\frac{m_{\tilde{i}}^5}{m^2_{3/2}M^2_\text{Pl}}\,,
\end{equation}
where $\tilde{i}=\tilde{g}, \tilde{f}$. The total MSSM contribution to the gravitino yield is $Y^\text{MSSM}_{3/2}=Y^\text{MSSM(sc)}_{3/2}+Y^\text{MSSM(dec)}_{3/2}$, and the relic density parameter reads
\begin{equation} \label{gravAbM}
\frac{\Omega^\text{MSSM}_{3/2}}{0.12\, h^{-2}}  \, \sim \,   \left[\frac{\hat{\gamma}_\text{sc}}{2} \left(\frac{m_{3/2}}{\text{GeV}}\right)  \left(\frac{T_\text{rh}}{10^{12} \text{GeV}}\right) +3\left(\frac{N}{46}\right)\left(\frac{10^2\text{GeV}}{m_{3/2}}\right) \left(\frac{m_{\tilde{f}_i}}{10^5\,\text{GeV}}\right)^3\right] +\frac{m_{3/2}}{m_\text{NLSP}} \frac{\Omega^{\text{(th)}}_\text{NLSP}}{0.12\, h^{-2}}.\, 
\end{equation}
The gravitino relic abundance sourced by the MSSM and messenger fields is illustrated in Fig. 2 and 3. In the case that the gravitino is the only sparticle with mass below the reheating temperature then the gravitino relic abundance  is given by a much different expression with dependence $\Omega_{3/2}\propto m^{-3}_{3/2} T^7_\text{rh}$ \cite{Benakli:2017whb, Dudas:2017rpa}.

Apart from particular cases, the above gravitino yield (\ref{gravAbM}) cannot be final because we neglected sources beyond the MSSM. The supersymmetry breaking sector is a necessary ingredient for all the consistent supersymmetric BSM scenarios \cite{Giudice:1998bp}. In general the extra fields only increase\footnote{It is though possible that the supersymmetry breaking sector leads to a suppressed $\Omega_{3/2}$, e.g due to $R$-symmetry restoration \cite{Dalianis:2011ic}, or a high temperature decoupling of the messenger fields \cite{Badziak:2015dyc}, or due to the dynamics of the sgoldstino field \cite{Co:2017pyf}, or due to feeble couplings in the supersymmetry breaking sector \cite{Tsao:2017vtn}.}
 the final $\Omega_{3/2}$, unless there is a late entropy production.  For example, thermalized messengers fields generically {\itshape equilibrate} the gravitinos  for  broad range of values of the Yukawa coupling at the messenger superpotential, $\lambda_\text{mess}\gtrsim 10^{-6}-10^{-5}$ \cite{Dalianis:2013pya}, and the relic gravitino density parameter reads
\begin{equation} \label{Oth1}
\frac{\Omega^{\text{(th)}}_{3/2}}{0.12\, h^{-2}}\, \sim \, 5 \times 10^{6} \left( \frac{m_{3/2}}{\text{GeV}} \right) \left[ {270}/{g_*\left(T^{\text{f.o.}}_{3/2}\right)} \right], 
\end{equation} 
where the freeze out temperature is here equal to the messenger mass scale, $T^{\text{f.o.}}_{3/2}\sim M_\text{mess}$. 
Even if $\lambda_\text{mess} \ll 1$ the thermal scatterings of messengers contribute to gravitino relic density with $\Omega^\text{mess}_{3/2}h^2 \sim 0.4 \left(M_\text{mess}/10^4\text{GeV}\right)\left(\text{GeV}/m_{3/2}\right)\left(m_{\tilde{g}}/\text{TeV}\right)^2$. 
In addition, the inflaton perturbative decay produces non-thermal gravitinos  with rate \cite{Kawasaki:2006hm, Endo:2007sz}
\begin{equation}
\Gamma(\Phi\rightarrow \tilde{G}\tilde{G}) \simeq \frac{|G^\text{(eff)}_\Phi|^2}{288\pi} \frac{m^5_\Phi}{m^2_{3/2} M^2_\text{Pl}}\,.
\end{equation}
Also gravitinos are produced during the preheating stage via its non-perturbative decay of the inflaton \cite{Giudice:1999am, Nilles:2001ry, Ellis:2015jpg, Ema:2016oxl, Hasegawa:2017hgd, Dalianis:2017okk, Addazi:2017ulg}, from the decay of the supersymmetry breaking field, see e.g. \cite{Ibe:2006rc, Hamaguchi:2009hy, Fukushima:2012ra}, or other moduli \cite{Endo:2006zj, Nakamura:2006uc, Dine:2006ii}.
Therefore, the estimation of the gravitino relic abundance based solely on the  MSSM sector gives a model independent albeit an underestimated and hence conservative value for the $\Omega_{3/2}$.

The $\Omega_{3/2}$ result could decrease in the case that extra fields interrupt the thermal phase, e.g. due to the domination of a non-thermal scalar field that produces entropy at low temperatures.  
Thanks to the dilution the gravitino cosmologically problematic supersymmetric scenarios may become viable possibilities. The tentative low entropy production is caused by the scalar $X$ that we do not identify and collectively call it {\itshape diluter}. We only assume that 
it interacts too weakly with the other fields, e.g. via gravitational interactions.
Therefore the gravitino relic density parameter is the {\itshape conditional} sum 
\begin{equation} \label{tot1}
\Omega^{\text{tot}}_{3/2} \simeq
\text{min}\left\{ \Omega^{\text{MSSM}}_{3/2} + \Omega^{\text{mess}}_{3/2}+ \Omega^{\text{inf}}_{3/2}+\Omega^\text{SB}_{3/2}\,,\,\,
\Omega^\text{(th)}_{3/2} \right\}
\end{equation} 
where $\Omega^{\text{MSSM}}_{3/2}$ is the contributions of the MSSM (scatterings and decays), $\Omega^{\text{mess}}_{3/2}$ is the contribution of messengers (scatterings and decays), $\Omega^{\text{inf}}_{3/2}$ is the contribution of the infationary perturbative and non-perturbative decay and $\Omega^\text{SB}_{3/2}$ is the contribution of the supersymmetry breaking field.   It is called conditional sum because the simple add of each  contribution may result in an overestimate of the gravitino abundance. For example the presence of thermalized messengers modifies the gravitino production from the MSSM sector \cite{Choi:1999xm, Dalianis:2013pya}. 
We mention that the sum (\ref{tot1}) is not strictly exact: 
it is well possible that contributions from non-thermal decays, that take place below the $T^\text{f}_{3/2}$ temperature, increase the $\Omega^\text{tot}_{3/2}$ beyond the $\Omega^\text{(th)}_{3/2}$ value. 

Finally, the presence of a scalar $X$ that produces low entropy modifies the result (\ref{tot1}) 
as will be discussed in the section 4. In such a case the density parameter (\ref{tot1}) value is renamed $\Omega^{<}_{3/2}$ in order to emphasize that it is sourced by processes taking place before the $X$ decay.

\begin{figure} \label{f3}
\centering
\begin{tabular}{cc} 
{(a)} \includegraphics [scale=.72, angle=0]{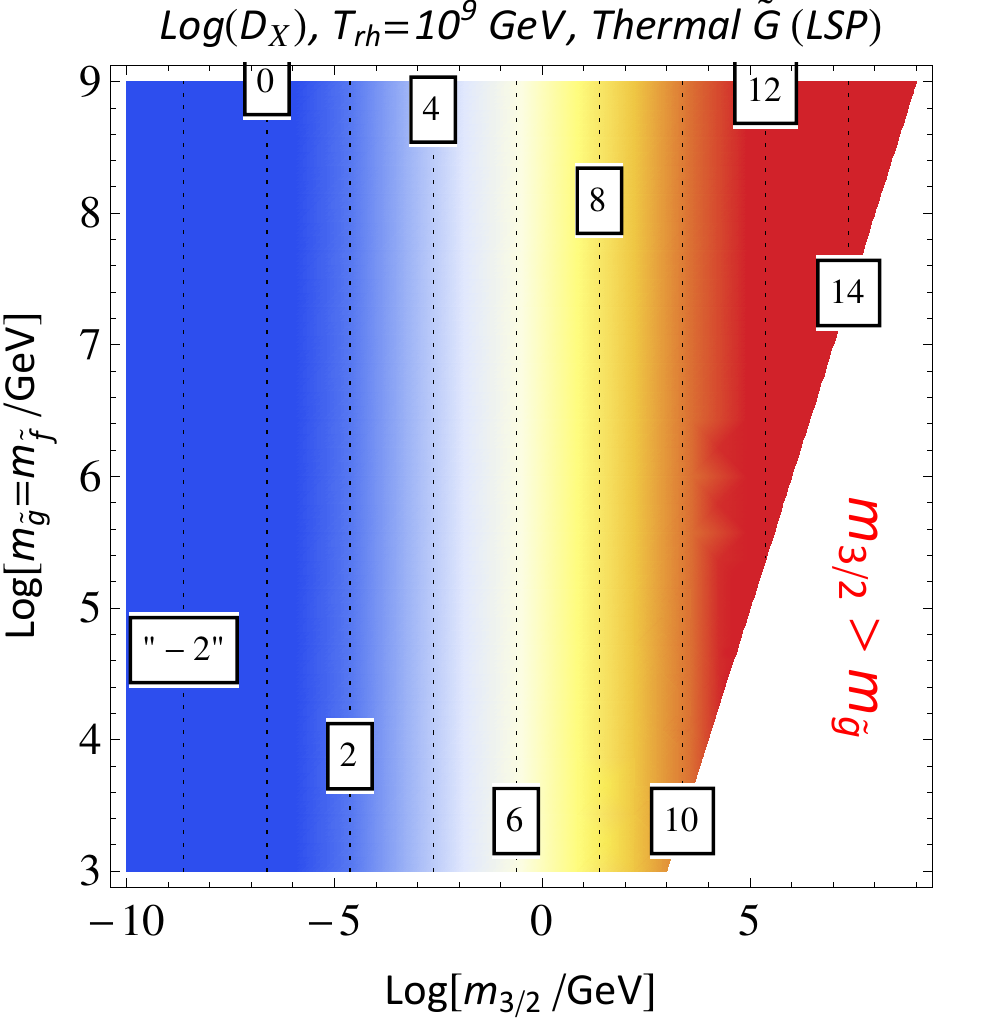} \quad
{(b)} \includegraphics [scale=.72, angle=0]{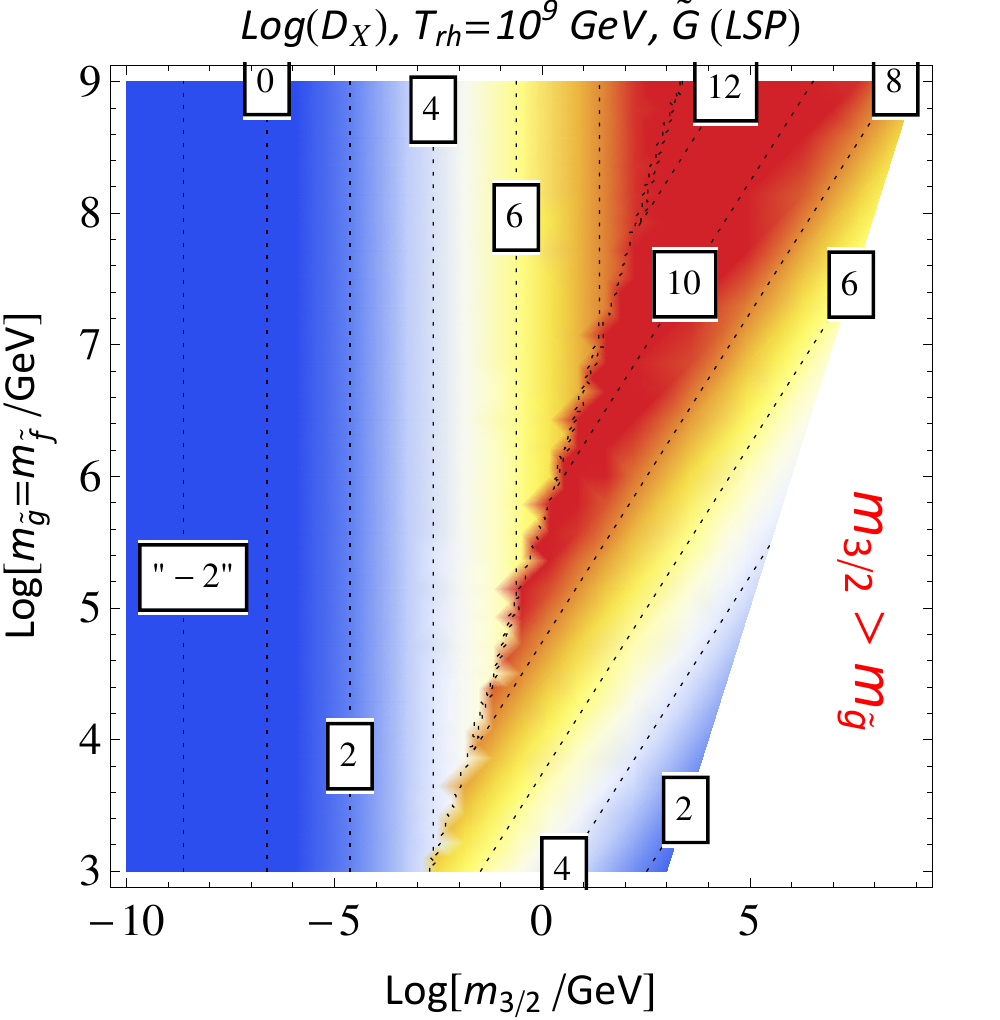}  \\
\end{tabular}
\caption{\small{Density and contour plot of the decadic logarithm of the {\itshape required} dilution for gravitino LSP and reheating temperature $T_\text{rh}=10^9$ GeV. In the {\itshape left panel} the gravitinos are thermal (heavy gravitinos can thermalize due to the messenger sector).  In the {\itshape right panel}  the contribution of messengers plus MSSM is considered, assuming a small enough messenger coupling so that gravitinos do not thermalize by the interactions with messengers, but only due to the MSSM. The messengers scale is taken to be $M_\text{mess}=10^8$ GeV. 
}}
\end{figure}

\subsection{Neutralino dark matter}

The lightest neutralino $\tilde{\chi}^0$ is the most representative example of a WIMP dark matter and an appealing candidate 
 thanks to its main merit: the thermal production mechanism. The thermal neutralino scenario however works best provided that the squark and slepton masses lie in the 50-100 GeV range \cite{Ellis:1983ew, Baer:1995nc} which has been excluded by collider searches. In addition the direct and indirect detection experiments shrink the parameter space of the neutralino with mass about the electroweak scale \cite{Baer:2011ab}.  Specific neutralino types, such as the higgsino, see e.g. \cite{Kowalska:2014hza}, or the annihilation mechanism can be invoked to match the $\Omega_{\tilde{\chi}^0} h^2$ to data, but in general a  rather heavy neutralino cannot be a viable thermal relic.

The neutralino $\tilde{\chi}^0$  decouples from the thermal bath at a freeze-out temperature $T^\text{f.o.}_{\tilde{\chi}^0}=m_{\tilde{\chi}^0}/x_f$, where $x_f\simeq 28 - \ln(m_{\tilde{\chi}^0}/\text{TeV}) + \ln(c/10^{-2})$, where $c/m^2_{\tilde{\chi}^0}= \left\langle  \sigma_{\tilde{\chi}^0} v  \right\rangle $ the non-relativistic $\tilde{\chi}^0$ annihilation cross section. In scenarios with split spectrum it is $c=3\times 10^{-3}$ for a mostly higgsino $\tilde{\chi}^0$ and $c=10^{-2}$ for mostly wino $\tilde{\chi}^0$ \cite{ArkaniHamed:2004yi}.  
If the reheating temperature is larger than $T^\text{f.o.}_{\tilde{\chi}^0}$ the neutralinos reach thermal and chemical equilibrium and the relic density parameter is UV insensitive and depends on the $\tilde{\chi}^0$ mass squared,  $\Omega^\text{(th)}_{\tilde{\chi}^0}\propto m^2_{\tilde{\chi}^0}$. When the sparticles masses lay well above the TeV scale the thermal neutralino scenario is disfavored  
and ususally non-thermal production scenarios are considered, e.g $\tilde{\chi}^0$ production via the decay of heavy gravitinos. 

The gravitinos, that are unstable, are produced via thermal scatterings, non-thermal decays of sfermions and possible decays of scalars beyond MSSM  such as the inflaton \cite{Kawasaki:2006hm, Endo:2007sz, Giudice:1999am, Nilles:2001ry, Ellis:2015jpg, Ema:2016oxl, Hasegawa:2017hgd, Dalianis:2017okk, Addazi:2017ulg} and the supersymmetry breaking field or other moduli \cite{Endo:2006zj, Nakamura:2006uc, Dine:2006ii}. Focusing on the MSSM sector the gravitinos dominate the universe either for large enough reheating temperature, $T_\text{rh} \gtrsim 5\times 10^{14}(m_{3/2}/10^5 \text{GeV})^{1/2}$ GeV, or large enough sfermion masses, $m_{\tilde{f}}\gtrsim 2 \times 10^{8}(m_{3/2}/10^5 \text{GeV})^{5/6}$ \cite{ArkaniHamed:2004yi}. The gravitinos decay when $\Gamma_{3/2}=H$ and the temperature after decay is
\begin{equation} \label{Tg}
T^\text{dec}_{3/2}= 6.8\,\left(\frac{m_{3/2}}{10^5\text{GeV}}\right)^{3/2} \left[\frac{10.75}{g_*(T^\text{dec}_{3/2})}\right]^{1/4}\, \text{MeV}\,.
\end{equation}
Apparently, it has to be $m_{3/2}>10^4$ GeV to avoid BBN complications \cite{Kawasaki:2004qu, Kawasaki:2008qe}. The $^4$He abundance implies that 
it must be $Y_{3/2} \lesssim  10^{-12},\quad \text{for}\quad  m_{3/2}=10-30$ TeV and for smaller $m_{3/2}$ values the bound becomes much severer, see e.g. \cite{Kawasaki:2004qu, Kawasaki:2006hm, Kawasaki:2008qe} for details. 
The gravitino decay populates the universe with neutralinos. 
Heavy enough gravitinos, $m_{3/2}\gg 10^7$ GeV, decay promptly so that $T^\text{dec}_{3/2} > T^\text{f.o.}_{\tilde{\chi}^0}$ and the neutralinos reach a thermal equilibrium. 
In the opposite case,  the neutralinos produced by the graviton decay are out of chemical equilibrium and either have a yield  $Y_{\tilde{\chi}^0}\sim Y_{3/2}$ for a radiation dominated universe, or $Y_{\tilde{\chi}^0} \simeq 3T^\text{dec}_{3/2}/(4{m_{3/2}})$ for a gravitino dominated early universe. Unless the reheating temperature is particularly high $T_\text{rh}>10^{14}$ GeV or the sfermions very massive, $m_{\tilde{f}} >10^8$ GeV the gravitinos do not dominate over the radiation, and the neutralino relic density parameter reads
\begin{equation} \label{th_nth}
\Omega^\text{}_{\tilde{\chi}^0} h^2 =\frac{m_{\tilde{\chi}^0}}{m_{3/2}} \left(\Omega^\text{MSSM(sc)}_{3/2} h^2+\Omega^{{\tilde{f}}\text{(dec)}}_{3/2}h^2 \right) + \Omega^\text{(th)}_{\tilde{\chi}^0} h^2  \quad\quad\quad \text{(Radiation-domination)}\,.
\end{equation}
Thus one finds
\begin{equation} \label{neuAb}
\frac{\Omega^\text{}_{\tilde{\chi}^0}}{0.12\, h^{-2}}\, \sim \,  \left( \frac{m_{\tilde{\chi}^0}}{\text{TeV}}\right)\left[\left( \frac{T_\text{rh}/2}{ 10^9 \text{GeV}}\right) + \left(\frac{10^5 \text{GeV}}{m_{3/2}} \right)^2  \sum_i g_i \left( \frac{m_{\tilde{f}_i}}{10^7 \text{GeV}}\right)^3 +\left(\frac{m_{\tilde{\chi}^0}}{\text{TeV}} \right) \left(\frac{10^{-3}}{c}\right) \right], 
\end{equation}
where $i$ runs up to $N=46$ for sfermions heavier than the gravitino
and $T^\text{dec}_{3/2} < T^\text{f.o.}_{\tilde{\chi}^0}$. 
The degrees of freedom at $T^\text{f.o.}_{\tilde{\chi}^0}$  were taken to be $g_*=86.25$. If gravitinos dominate the early universe that is $D_{3/2}\gg 1$, then the relic density parameter of the non-thermally produced neutralinos is
\begin{equation} \label{GD}
\frac{\Omega^\text{(n-th)}_{\tilde{\chi}^0}}{0.12\, h^{-2}} \sim  10^5 \left(\frac{m_{\tilde{\chi}^0}}{\text{TeV}}\right) \left(\frac{m_{3/2}}{10^5\,\text{GeV}} \right)^{1/2}\quad\quad\quad (\tilde{G}- \text{domination})\,.
\end{equation}
It is possible that the non-thermally produced neutralinos from the gravitino decay  achieve a chemical equilibrium for $n_{\tilde{\chi}^0} \left\langle \sigma_{\tilde{\chi}^0} v \right\rangle > H(T^{\text{dec}}_{3/2})$. It is $ \left\langle  \sigma_{\tilde{\chi}^0} v  \right\rangle \propto 1/m^2_{\tilde{\chi}^0}$ and $H(T^\text{dec}_{3/2}) \propto m^3_{3/2}$, hence for a wino-like neutralino at the TeV scale and $m_{3/2}>10^5$ GeV pair-annihilation can take place \cite{Moroi:1999zb}. The neutralinos annihilate until their number density becomes $n^\text{crit}_{\tilde{\chi}^0} \sim 3H/\left\langle \sigma_{\tilde{\chi}^0} v \right\rangle$ and the relic density parameter is for this case, 
$\Omega^\text{(ann)}_{\tilde{\chi}^0}  = \Omega^\text{(th)}_{\tilde{\chi}^0}   \left ({T^\text{f.o.}_{\tilde{\chi}^0}}/{T^\text{dec}_{3/2}}\right)$,
 that is enhanced by the ratio $(T^\text{f.o.}_{\tilde{\chi}^0}/T_{3/2})$ compared to the thermal abundance. This is an appealing scenario, called {\it annihilation scenario}, because the critical value $n^\text{crit}_{\tilde{\chi}^0}$ behaves as an attractor and determines the relic abundance of neutralino (mostly wino) LSP, making it independent of the primordial gravitino relic abundance \cite{Moroi:1999zb}. Nevertheless it hardly works when one departs from the TeV scale neutralino. It is also much constrained from the indirect detection experiments.

\begin{figure} \label{f4}
\centering
\begin{tabular}{cc} 
{(a)} \includegraphics [scale=.72, angle=0]{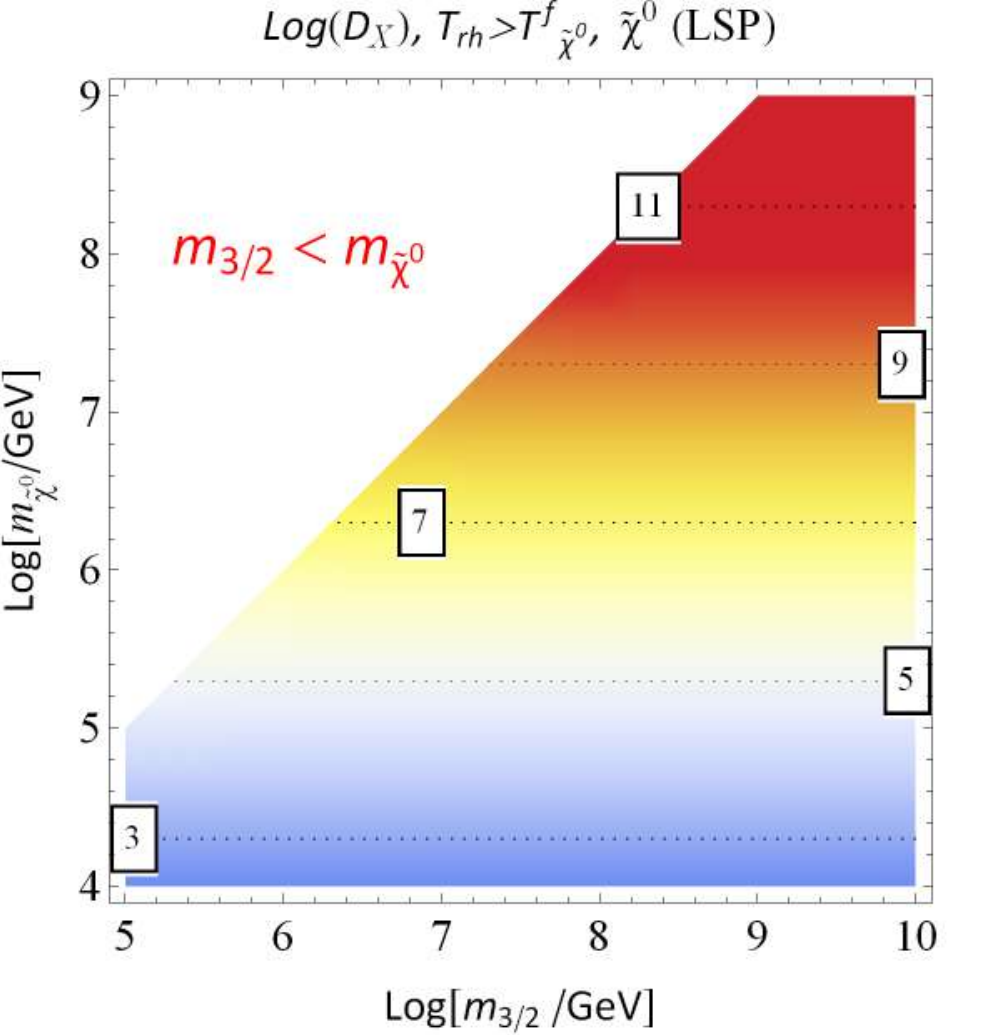} \quad
{(b)} \includegraphics [scale=.54, angle=0]{NnewF.pdf}  \\
\end{tabular}
\caption{\small{ Density and contour plot of the decadic logarithm of the {\itshape required} dilution 
for neutralino stable LSP. 
In the left panel the neutralino abundance is the thermal one. In the right the reheating temperature is $T_\text{rh}=10^{12}$ GeV the sfermions are $10^3$ times heavier than the gravitinos and hence the neutralino yield is dominated by the decay of gravitinos produced from sfermion decays; in the right bottom corner of the plot the neutralinos thermalize due to large $T^\text{dec}_{3/2}$. The gravitino mass is taken to be $m_{3/2}>10^5$ GeV to avoid BBN constraints.
}}
\end{figure}

In the section 4 the non-thermal scenario, that is often-called {\it branching ratio} scenario, where the $\tilde{\chi}^0$ is produced non-thermally during the low entropy production caused by the diluter $X$ field will be discussed, and in section 5 we will consider the production of $\tilde{\chi}^0$  from the supersymmetry breaking field aiming at a complete analysis.


\subsection{Axino dark matter}

In the sake of completeness of the basic LSP scenarios, we briefly comment here on the axino dark matter. In supersymmetry, the axion solution to the strong CP problem comes with an extra scalar, the saxion and a fermion, the axino $\tilde{a}$. If the axino is the LSP it is a well motivated dark matter candidate \cite{Covi:1999ty, Choi:2011yf}. It freezes out at high temperatures $T^\text{f.o.}_{\tilde{a}}\sim 10^{11}\text{GeV}(f_a/10^{12} \text{GeV})^2$, where $f_a$ the axion decay constant.
At lower temperatures it can be produced from thermal scatterings and decays. 
In that case, for a radiation dominated universe, the axino relic density parameter is the sum of the contributions from thermal scatterings, the gravitino decay and the NLSP decays 
\begin{equation} \label{n-th-a}
\Omega^\text{}_{\tilde{a}}  \simeq  \frac{m_{\tilde{a}}}{m_{3/2}} \left(\Omega^\text{MSSM(sc)}_{3/2} +\Omega^{{\tilde{f}}\text{(dec)}}_{3/2} \right)+\frac{m_{\tilde{a}}}{m_\text{NLSP}}\Omega^\text{}_\text{NLSP}   + \Omega^\text{MSSM(sc)}_{\tilde{a}}\,, 
\end{equation}
for $T^\text{dec}_{3/2}$ below the NLSP freeze out temperature. We note that the two body decay of a squark to an axino is subdominant for gluino masses less than squark mass \cite{Covi:2002vw}. It is $\Omega^\text{MSSM(sc)}_{\tilde{a}} \sim 2.8 \times 10^8 (m_{\tilde{a}}/\text{GeV}) Y_{\tilde{a}}$ where $Y_{\tilde{a}}(\text{KSVZ})\sim 10^{-7} (T_\text{rh}/10^4 \text{GeV})(10^{11}\text{GeV}/f_a)^2$ for the KSVZ axion model, see e.g \cite{Brandenburg:2004du}, and $Y_{\tilde{a}}(\text{DFSZ})\sim 10^{-5} (\mu /\text{TeV})^2 (10^{11}\text{GeV}/f_a)^2$ for the DFSZ axion model where $\mu$ the superpotential Higgs/Higgsino parameter, see e.g \cite{Bae:2011iw}. 

For axino mass not much smaller than the NLSP, the axino dark matter case is quite similar to the neutralino LSP. For $m_{\tilde{a}}\gtrsim$ TeV the axino dark matter is also cosmologically problematic since its relic density parameter generally violates the $\Omega_\text{DM}h^2=0.12$ bound, and the essential conclusion is that, in general, a special thermal history of the universe is required for the axino dark matter scenario as well. Remarkably in these models, the saxion can play the r\^ole of the diluter $X$ for its condensate decay can produce late entropy that successfully decreases the LSP abundance \cite{Kawasaki:2008jc}, see also \cite{Co:2016fln} for some recent results on the reheating temperature and the $\Omega_\text{LSP}$ constraint.


\section{Alternative cosmic histories and supersymmetry} 

The overview of the predicted relic density of supersymmetric dark matter in section 3 suggests that the  observational value of $\Omega_\text{DM}h^2$ gets generally severely violated when the sparticle masses increase. For gravitino and neutralino LSP one can collectively write down a general scaling with respect to the mass parameters and temperature
\begin{equation} \label{g1}
\Omega^\text{}_{3/2} \propto \, m^\alpha_{3/2} \, \left(\frac{m_{\tilde{g}}}{m_{3/2}}\right)^\beta \, \left(\frac{m_{\tilde{f}}}{m_{3/2}}\right)^\gamma\, T^\delta_\text{rh} \,\,,\quad\quad\quad\, m_{3/2}<m_{\tilde{g}},m_{\tilde{f}}\,,
\end{equation}
and
\begin{equation} \label{g2}
\Omega^\text{}_{\tilde{\chi}^0} \propto \, m^{\tilde{\alpha}}_{\tilde{\chi}^0} \,\, m_{3/2}^{\tilde{\beta}} \, \left(\frac{m_{\tilde{f}}}{m_{3/2}}\right)^{\tilde{\gamma}}\, T^{\tilde{\delta}}_\text{rh}\,\,,\quad\quad\quad\,\,\quad\quad\, m_{\tilde{\chi}^0}<m_{3/2},m_{\tilde{f}}
\end{equation}
where the exponents $(\alpha, \beta, \gamma, \delta)$ and $(\tilde{\alpha}, \tilde{\beta}, \tilde{\gamma}, \tilde{\delta})$ are either positive or zero, depending on the dark matter production mechanism considered. 

The predicted supersymmetric dark matter overdensity for "unnatural" supersymmetry can be reconciled with the $\Omega_\text{DM}h^2$ bound if the reheating temperature is rather low or late entropy production takes place. Remarkably, both solutions imply that an alternative cosmic history takes place if supersymmetry is a symmetry of nature. By the term {\it alternative cosmic history} we mean that the radiation domination phase after inflation was interrupted or delayed by a cosmic era, where a fluid $X$ with barotropic parameter $w_X <1/3$ dominated the energy density of the early universe. As discussed in the introduction and in section 2 such a cosmic era impacts the observable values $n_s$ and $r$. 
In order to quantify this effect we consider in our analysis below different cosmic histories and different supersymmetry breaking schemes.
We follow the base line framework of the benchmark supersymmetry breaking scenarios with
either gravitino or neutralino LSP and degenerate or split mass spectrum. 
The scale of supersymmetry breaking, represented by the general sfermion mass $\tilde{m}$, is taken to be from the TeV scale up to the energy scale of the reheating temperature.

\subsection{Low reheating temperature} 

The reheating temperature of the universe after inflation can be rather low if the inflaton decay rate, $\Gamma_\text{inf}$, is small enough or if it is the result of the decay of a weakly coupled scalar unrelated to the inflaton\footnote {Low reheating temperatures may be caused by a scalar field $X$ (or more than one scalar) with relatively long lifetime, $\Gamma_X \ll \Gamma_\text{inf}$ that dominated the energy density of the universe before the inflaton decay, e.g if the $X$ is frozen during inflaton oscillations with the $\rho_X\sim m^2_X X^2$ sufficiently large that sources some extra e-folds of $X$ inflation. In this case, the reheating temperature at the expressions (\ref{g1}) and (\ref{g2}) is $T_\text{rh} = T^\text{dec}_X$.}. 
 In this case, the dark matter production due to processes sensitive to the maximum temperature gets suppressed.


We call {\it low} reheating temperature scenarios those with  $T_\text{rh}\lesssim 10^5$ GeV. For gravitino LSP 
the yield from thermal scatterings decreases when the reheating temperature decreases, and the NLSP-decays to gravitinos account for the leading contribution to $\Omega_{3/2}$ for $m_{3/2} \sim m_\text{NLSP}$. 
On the other hand, for neutralino LSP  the UV-sensitivity of the $\Omega_{\tilde{\chi}^0}$ to processes that take place at high temperatures is small. 
The neutralino abundance is IR-sensitive and it is mostly determined at the freeze out temperature $T^{\text{f.o.}}_{\tilde{\chi}^0}$.

Both for gravitino and neutralino LSP, the observational bound  $\Omega_\text{DM}h^2=0.12$ is generally violated for $m_\text{LSP}> {\cal O}(\text{TeV})$ and $\tilde{m}<T_\text{rh}$. 
Apparently, in the MSSM the  $\Omega_\text{DM}h^2=0.12$ can be satisfied for $m_\text{LSP} \lesssim {\cal O}(\text{TeV})$   or for the particular case that $\tilde{m}\sim T_\text{rh}$ where the Boltzmann suppression may play a critical r\^ole. 
Note that the $X$ domination cosmic phase is a decaying particle dominated phase, hence entropy is gradually produced for $\Gamma_X /H<1$, where $\Gamma_X$ the decay rate of the $X$ particle. The maximum reheating temperature is greater that $T_X^\text{dec}$ and this has implications for the relic LSP density   \cite{Giudice:2000ex}. 
If the LSPs reach chemical equilibrium before reheating,  the relic LSP energy density, is roughly given by $\Omega^\text{(th)}_{\text{LSP}}\times T^3_\text{rh}T^\text{f.o.}_{\text{LSP}}/(T^\text{f.o., new}_{\text{LSP}})^4$ where  $T^\text{f.o., new}_{\text{LSP}}$ and $T^\text{f.o.}_{\text{LSP}}$  are the freeze-out temperatures for $T_\text{rh}\ll m_\text{LSP}$ and  $T_\text{rh}\gtrsim m_\text{LSP}$ respectively   \cite{Giudice:2000ex}. 
We have also called $T_\text{rh}$ the $X$ decay temperature, $T_X^\text{dec}$.
On the other hand, if $T_\text{rh}\ll m_\text{LSP}$ and the LSPs never reach a chemical equilibrium then the relic density has a dependence $\Omega_\text{LSP} \propto T^7_\text{rh}$ \cite{Chung:1998rq}. Finally, if the LSPs are produced non-thermally from the $X$ decay and reach chemical equilibrium then the relic density reads $\Omega^\text{(th)}_{\text{LSP}} \times({T^\text{f.o.}_{\text{LSP}}}/{T_\text{rh}})$, see \cite{Gelmini:2006pw} for a brief overview on the topic. Note that these scenarios that can reconcile heavy supersymmetry with the observational bound $\Omega_\text{DM}h^2=0.12$ work mainly for the TeV neutralino  dark matter scenario and share the common feature that $T_\text{rh}<T^\text{f.o.}_{\text{LSP}}$.

To this end, one draws the general conclusion that the gravitino or neutralino LSP relic density  for "unnatural" supersymmetry $\tilde{m}>{\cal O}(\text{TeV})$ and  $m_\text{LSP}> {\cal O}(\text{TeV})$ requires the reheating temperature after inflation to be below or about the supersymmetry breaking scale,
\begin{equation} \label{Tl}
T_\text{rh} \lesssim \tilde{m}\,.
\end{equation}
Otherwise the dark matter is overabundant. Let us mention that the very interesting scenario of EeV gravitino \cite{Benakli:2017whb, Dudas:2017rpa, Dudas:2017kfz} falls into this category, although there the reheating temperature is not low. Remarkably the relation (\ref{Tl}) implies that, if there no late entropy production, {\it the measurement of the reheating temperature via the $(n_s, r)$ values indicates a lower bound for the supersymmetry breaking scale}.
Interestingly enough, the opposite limit $\tilde{m}\sim {\cal O}(\text{TeV})$ and  $m_\text{LSP} \lesssim {\cal O}(\text{TeV})$  is in the probing range of terrestrial colliders and detection experiments, a fact that manifests the complementarity of the cosmological investigation.

\subsection{Late entropy production}
If the inflaton decay reheated the universe it is generally expected that $T_\text{rh}\gtrsim 10^9$ GeV, for inflaton mass $m_\Phi \sim 10^{13}$ GeV.
Such reheating temperatures mean that the dark matter generation processes that take place at high temperatures are critical for the determination of the dark matter abundance.  

The abundance that is more sensitive to UV processes is that of the gravitino. In the MSSM framework, the leading contribution to $Y_{3/2}$ depends on the maximum temperature after reheating and the decays of the heavy thermalized sfermions. As illustrated in the Fig. 2 and 3, the $\Omega_{3/2}h^2/0.12$ increases with the $\tilde{m}$ and $m_{3/2}$. For $T^\text{f.o.}_{3/2}<T_\text{rh}$  the gravitino abundance is the thermal one, and in particular cases it may be enhanced due to late sfermion decays.

In the neutralino LSP case, the $\tilde{\chi}^0$ thermal abundance freezes out at the temperature $T^\text{f.o.}_{\tilde{\chi}^0}$ and the $Y_{\tilde{\chi}^0}$ receives extra contributions from the gravitino late decays. If the gravitino is heavy enough it can be $T^\text{dec}_{3/2}> T^\text{f.o.}_{\tilde{\chi}^0}$ and the neutralinos from the gravitino decay equilibrate. Generally the neutralino abundance increases as $m^2_{\tilde{\chi}^0}$ and the relic neutralino density is too large for $\tilde{m}$ and $m_{\tilde{\chi}^0}$ beyond the TeV scale, see Fig. 4. Moreover TeV scale supersymmetry with neutralino LSP, although compatible with the $\Omega_\text{DM}$ bound, is disfavoured for reheating temperatures $T_\text{rh}\gtrsim 10^8$ GeV due to BBN constraints on the late decaying gravitino abundance \cite{Kawasaki:2004qu,Kawasaki:2008qe}.  

One concludes that scenarios with reheating temperatures  $T_\text{rh}>\tilde{m}$ and $m_\text{LSP} >{\cal O}$(TeV) are compatible only if late entropy production takes place. The above remarks are synopsised in the following conditions, 
\begin{itemize} \raggedright
	\item \quad $\text{If} \quad\quad\quad  D_X = 1 \quad\quad\quad \text{then} \quad\quad\quad   T_\text{rh}\lesssim \tilde{m} \quad\quad \text{or}  \quad\quad \tilde{m}\sim \text{TeV}$ \hfill{\textbf{(A)}}
	\item \quad $ \text{If} \quad \quad\quad {\cal O}(\text{TeV})\, < \,  (m_\text{LSP}\,, \tilde{m}) <\, T_\text{rh}    \quad\quad\quad \text{then} \quad\quad\quad  D_X \neq 1 \,, \quad\quad\quad\quad $ \hfill{\textbf{(B)}}
\end{itemize}
where $\tilde{m}$ the sparticle mass scale.

Hence, scenarios with high reheating temperature generally require an extra scalar field that causes dilution.

\subsection{The diluter field $X$}

In supersymmetric theories  generically exist scalar fields with rather flat potentials and very weak or $M_\text{Pl}$ suppressed interactions. 
These kind of scalars, that are common in supergravity and superstring theories,
are here collectively labeled $X$. The $X$ domination, either due to its nearly constant potential energy or due to the energy stored in its oscillations about the vacuum, 
dilutes the LSP abundance $D_X$ times and supplements it with the contribution from the diluter decay
\begin{equation} \label{dil}
\Omega^{<}_\text{LSP} \rightarrow \frac{\Omega^{<}_\text{LSP}}{D_X} + \Omega^{X}_\text{LSP}\, \equiv \, \Omega^{\text{}}_\text{LSP}\,,
\end{equation}
where we labeled $\Omega^{<}_\text{LSP}$ the LSP abundance {\it before} the $X$ decay. 
In order to specify the $\Omega^{\text{}}_\text{LSP}$ the system of the three interacting cosmic fluids of $X$, LSP and radiation has to be solved and we refer the reader to references \cite{McDonald:1989jd, Chung:1998rq, Giudice:2000ex} for detailed analytic results.
For gravitino or axino LSP the above expression generally applies. For the neutralino LSP one should also check whether the conditions  (i) $T^\text{dec}_X<T^\text{f.o.}_{\tilde{\chi}^0}$ and (ii) $n_{\tilde{\chi}^0}\left\langle \sigma v \right\rangle < H(T^\text{dec}_X)$ hold. If not, then in the case (i) the neutralinos might reach a thermal equilibrium value $Y^\text{(th)}_{\tilde{\chi}^0}$. In the case (ii) pair annihilations take place until the neutralino yield reaches the value $Y^\text{(th)}_{\tilde{\chi}^0}\times(T^\text{f.o.}_{\tilde{\chi}^0}/T^\text{dec}_X)$; this corresponds to the so-called annihilation scenario and works mostly for wino-like LSP with TeV  mass scale.  Let us mention here that the radiation produced from the decay of the $X$ particles for the times $\Gamma_X/H<1$ can produce neutralinos even for $T^\text{dec}_X<T^\text{f.o.}_{\tilde{\chi}^0}$ \cite{McDonald:1989jd,  Giudice:2000ex}, which accounts for an extra contribution to $\Omega^{\text{}}_\text{LSP}$ that may be important in particular scenarios without, however, modifying the conclusions of the current analysis.  
Finally, the $\Omega^{X}_\text{LSP}$ depends on the branching ratio Br$^X_\text{LSP}$ of the diluter into two LSPs (directly or via cascade decays) and the $X$ decay temperature $T^\text{dec}_X$. The LSP yield from the $X$ decay reads
\begin{equation}
Y^{X}_\text{LSP} \equiv \frac{n_\text{LSP}}{s} =\frac32 \,\text{Br}^X_\text{LSP} \, \frac{T^\text{dec}_X}{m_X}\,.
\end{equation}
If the $Y^{X}_\text{LSP}$ is subdominant  the observed dark matter has to be produced by processes taking place at higher temperatures than $T^\text{dec}_X$ and was appropriately  diluted by the decay of the scalar $X$. On the other hand, if the dilution $D_X$ decreases the initial LSP abundance to negligible levels, then the LSP production from the $X$ decay should fit the observed dark matter abundance. 
The constraint $\Omega^{\text{}}_\text{LSP}h^2 \leq 0.12$ implies
\begin{equation} \label{Dmin}
D_X \geq D_X^\text{min} \equiv \frac{\Omega^{<}_\text{LSP}}{0.12 \,h^{-2}}\,,
\end{equation}
which determines the dilution magnitude and consequently the shift in the spectral index (\ref{Dn}). The $D^\text{min}_X$ is referred as the {\itshape required} dilution throughout the text, necessary to give at most a critical density of LSP particles today.

The $X$ decay is not free from constraints. 
It must decay before the BBN \cite{Kawasaki:2004qu}, not overproduce LSPs and not overproduce late decaying particles such as gravitinos. 
In the simple but quite unnatural case that the $X$ is lighter than LSP then it is Br$^X_\text{LSP}=0$ and the $X$ decay generates Standard Model radiation only. The Br$^X_\text{LSP}=0$ scenario becomes natural if $m_\text{LSP}<m_X<2m_\text{LSP}$ since the channel $X\rightarrow \tilde{G}\tilde{G}$ or $\tilde{\chi}^0\tilde{\chi}^0$ is forbidden due to kinematic constraints.

If the decay of the $X$ produces LSPs or other late decaying particles the relevant branching ratios have to be considered. 
This is a model dependent issue and should be examined in the context of each model. In the next section we consider the supergravity $R^2$ inflation and we take into account the $X$ decay rate and channels.
Actually, the details of the $X$ decay do not change any of the conclusions synopsised in the conditions (A) and (B). The minimum amount of dilution (\ref{Dmin}) is {\itshape necessary} regardless the diluter branching ratios, and this is a key point of this work.

\begin{figure} \label{dilXcor}
\centering
\begin{tabular}{c} 
{} \includegraphics [scale=.67, angle=0]{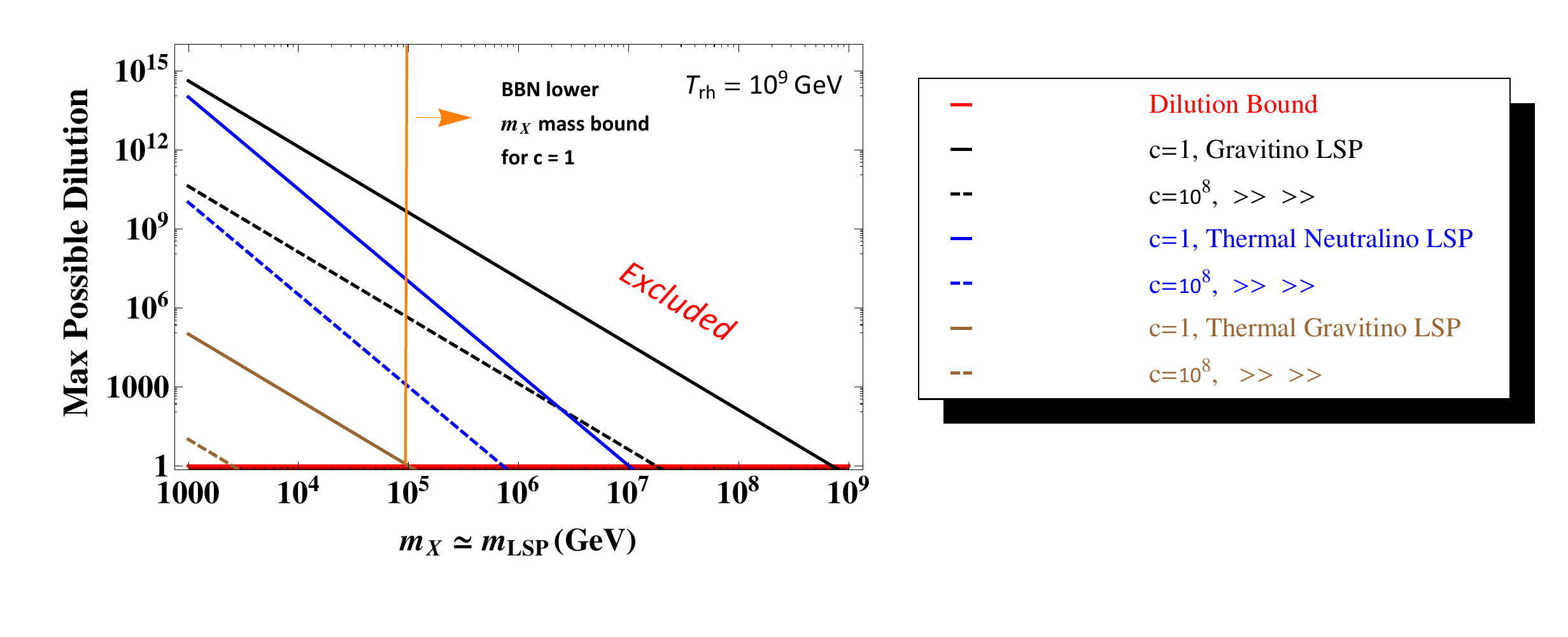} 
\end{tabular}

\caption {\small {The maximum possible dilution size, caused by a scalar $X$ condensate,  with respect to the LSP mass, for gravitino LSP (black, brown) and neutralino LSP (blue).   We have made the conservative assumption $m_X \simeq m_\text{LSP}$ that maximizes the diluter $X$ lifetime. In the area above the lines it is $\Omega^{<}_\text{LSP}h^2/D_X > 0.12$, hence it is an {\it excluded} parameter area. The solid and dashed lines correspond to $c=1$ and $c=10^8$ according to the parametrization (\ref{Xdec}). For a gravitationally decaying diluter (c=1), the thermal gravitino scenario (brown solid line) is excluded because the $X$ spoils the BBN predictions. 
 The plot demonstrates the decrease of the dilution efficiency for large supersymmetry breaking scale and $T_\text{rh}=10^9$ GeV.
}}
\end{figure}

\subsection{The maximum possible dilution due to a scalar condensate} 

If the diluter mass is about or larger than the LSP mass, $m_X \gtrsim m_\text{LSP}$, then the dilution magnitude is {\itshape correlated} with the supersymmetry breaking scale.  A late time entropy production takes place when the radiation dominated era gets interrupted by an $X$ domiation era at $T^\text{dom}_X<T_\text{rh}$, where $T_\text{rh}$ is the reheating temperature caused by the inflaton decay. For an oscillating scalar field the dilution magnitude is $D_X\simeq T^\text{dom}_X/T^\text{dec}_X$. The decay rate of the $X$ scalar can be parametrized as
\begin{equation} \label{Xdec}
\Gamma^\text{}_X = \frac{c}{4\pi} \frac{m^3_X}{M^2_\text{Pl}}\,,
\end{equation} 
and the $X$ decay temperature is $T^\text{dec}_X \simeq (\pi^2 g_*/90)^{-1/4}(\Gamma^\text{}_X M_\text{Pl})^{1/2}$. For $c\sim1$ the $X$ decays gravitationally and $T^\text{dec}_X \sim 4 \,\text{MeV}\, (M_X/10^5\text{GeV})^{3/2} $.
 For $c\gg 1$ non-gravitational decay channels exist; for example if the $X$ field has Yukawa-like coupling $y_X$ to light degrees of freedom then it is $\Gamma_X=y^2_X m_X/8\pi$.
For the borderline case that $T^\text{dom}_X=T_\text{rh}$ and $m_X=m_\text{LSP}$  the dilution magnitude due to an oscillating scalar field, $D_X$, reaches a maximum 
value. 
Consequently, a minimum  
value for the $\Omega^{<}_\text{LSP}h^2/D_X$ exists  which obviously must be below the observational value $\Omega_\text{DM}h^2=0.12$. 

In particular, for gravitino LSP the lowest $T^\text{dec}_X$ value is achieved for $\Gamma^\text{min}_X = ({c}/{4\pi}) {m^3_\text{LSP}}/{M^2_\text{Pl}}$ and $c\sim 1$. Assuming that gravitinos are mainly produced by thermal scatterings then the maximum possible dilution value, $D^\text{max}_X=T_\text{rh}/T^\text{dec(min)}_X\geq D_X$, yields the lower bound
\begin{equation} \label{c1}
\, \hat{\gamma}_\text{sc} \,c^{1/2}\, \left( \frac{m_{3/2}}{7\times 10^8 \,\text{GeV}}  \right)^{5/2}\, <  \, \frac{\Omega^{<}_{3/2} h^2/0.12}{D^\text{}_X} \, \leq \, 1\,,
\end{equation}
where $\Omega^{<}_{3/2}=\Omega^\text{MSSM(sc)}_{3/2}$, $\hat{\gamma}_\text{sc} \gtrsim1$, see Eq. (\ref{Grsc}), and the parameter $c$ is explicitly written. Note that although the $T_\text{rh}$ is dropped out in the above relation  it must be $T_\text{rh}> m_{3/2}$. The constraint (\ref{c1}) says that the abundance of gravitino LSPs produced from thermal scatterings in the plasma is possible to get diluted to observationally acceptable values by an oscillating scalar field that obtains mass from the supersymmetry breaking {\it only if}
\begin{equation}
m_{3/2} < 7\times 10^8\,\text{GeV}\,.
\end{equation}
The constraint becomes more severe if $\hat{\gamma}_\text{sc} \gg 1$, that is, if $m^2_{3/2} \ll m^2_{\tilde{g}} <T^2_\text{rh}$, or for a non-gravitational scalar $X$, $c\gg 1$ or for $m_X \gg m_{3/2}$. 
For thermalized gravitinos instead, the formula (\ref{Oth1}) applies and the maximum possible dilution magnitude gives the following constraint
\begin{equation} \label{c2}
c^{1/2}\, \left( \frac{m_{3/2}}{ 10^5 \,\text{GeV}}  \right)^{5/2} \left( \frac{10^9\,\text{GeV}}{T_\text{rh}} \right)\, <  \, \frac{\Omega^{<}_{3/2} h^2/0.12}{D^\text{}_X} \, \leq \, 1\,,
\end{equation}
where $\Omega^{<}_{3/2}=\Omega^\text{eq}_{3/2}$. We see from (\ref{c2}) that typical reheating temperatures $T_\text{rh}=10^9-10^{12}$ GeV imply a mass bound $m_{3/2}\lesssim 10^6$ GeV for thermalized LSP gravitinos. Although such heavy gravitinos hardly get thermalized via interactions with the MSSM plasma, thermalized messengers can bring them to thermal equilibrium. Again here, the bound (\ref{c2}) becomes more severe for $c\gg 1$ or for $m_X \gg m_{3/2}$.

When the neutralino is the LSP the ${\tilde{\chi}^0}$ relic abundance is determined at the freeze out temperature that is $T^{\text{f.o.}}_{\tilde{\chi}^0}\sim m_{\tilde{\chi}^0}/20$. If the decay temperature of the $X$ field is below the $T^{\text{f.o.}}_{\tilde{\chi}^0}$ the neutralinos number density will get diluted. 
For thermally produced neutralinos the maximum possible dilution magnitude, for $T^\text{dom}_X=T_\text{rh}$ and $m_X\sim m_{\tilde{\chi}^0}$, gives the constraint
\begin{equation} \label{c3}
c^{1/2}\, \left( \frac{m_{\tilde{\chi}^0}}{ 10^7 \,\text{GeV}}  \right)^{7/2} \left( \frac{10^9\,\text{GeV}}{T_\text{rh}} \right)\, <  \, \frac{\Omega^{<}_{\tilde{\chi}^0} h^2/0.12}{D^\text{}_X} \, \leq \, 1\,.
\end{equation}
Thus, neutralino masses $m_{\tilde{\chi}^0}> 10^7\text{GeV}$ for $T_\text{rh}\lesssim 10^9$GeV cannot be reconciled with cosmological scenarios where an oscillating scalar field dilutes the thermal plasma.
If the leading contribution to $\Omega^{<}_{\tilde{\chi}^0}$ comes from the decays of gravitinos which are respectively produced from sfermion decays for $m_{\tilde{f}} > m_{3/2} > m_{\tilde{\chi}^0}$ 
the expression (\ref{GD})  has to be used and another constraint for the $\tilde{\chi}^0$ mass 
is obtained.

This correlation among the dilution size due to scalar oscillations and the $m_\text{LSP}$ (or alternatively the supersymmetry breaking scale) is an {\itshape extra} and important constraint on these scenarios, see Fig. 5.
The constraints on the LSP mass can be raised if {\it thermal inflation} takes place and then the dilution size is given by the expression (\ref{FD}). In such a case it is $c\gg 1$ since the $X$ is not a gravitationally decaying scalar due to the necessary presence of Yukawa couplings of $X$ with the thermalized degrees of freedom, that regulate the decay rate. The gravitational diluter scenario is certainly a less model dependent and a more generic one.


\section{A concrete example: The $R+R^2$ (super)gravity inflationary model}

Inflation is the leading paradigm for explaining the origin of the primordial density perturbations 
that grew into the CMB anisotropies. If the early Universe is described by a typical model of
inflation that naturally explains the statistical properties of the density and the expected tensor perturbations then the precision $(n_s, r)$ measurement can give us physical evidences for the radiation dominated era before the epoch of nucleosynthesis.

In this section we will apply the previously obtained results on the $R^2$ gravity and supergravity inflation models in order to perform a full estimation of the theoretically expected values for the $n_s$ and $r$ observables. The Starobinsky $R^2$ inflation model is particularly motivated because it is placed in the center of the likelihood contour, nonetheless it is self evident that a similar analysis can be performed for any other inflation model. In the following, preliminaries of the $R^2$ gravity and supergravity inflation models will be reviewed.  For the supergravity $R^2$ model new predictions for the $(n_s, r)$ observables will be derived, depending on the supersymetry breaking scheme, and the phenomenology of the two models will be compared.

\subsection{The Starobinsky $R^2$ inflation}

The  Starobinsky model \cite{Starobinsky:1980te} is an $f(R)$ gravity model described by the Lagrangian
\begin{equation} \label{star}
e^{-1} {\cal L} = -\frac{M_\text{Pl}^2}{2} R+ \frac{M_\text{Pl}^2}{12 m^2} R^2  \, . 
\end{equation}
This theory is conformally equivalent to the Einstein gravity with a scalar field $\varphi$, the scalaron,  minimally coupled to gravity
\begin{equation}
\label{effS}
e^{-1} {\cal L} =  - \frac{M_\text{Pl}^2}{2} R 
- \frac12 \partial \varphi \partial \varphi  - \frac34 m^2 M_\text{Pl}^2 \left( 1  -  e^{- \sqrt \frac23 \varphi/M_\text{Pl}}  \right)^2   \,.
\end{equation}
From the CMB normalization \cite{Ade:2015xua} we get $m \simeq 1.3 \times 10^{-5} M_\text{Pl}$.
The inflationary predictions of the $R^2$ theory \cite{Mukhanov:1981xt} at leading order are given by the following expressions of the primordial spectra and tensor-to-scalar-ratio $r_*= 16\epsilon_*$,
\begin{equation} \label{IR1}
n_s=1-\frac{2}{N_*} \,, \quad\quad \frac{d n_s}{d\ln k} \simeq -\frac{2}{N^2_*}\,, \quad\quad r_*= \frac{12}{N^2_*}\,.
\end{equation}
Also, the tensor spectral tilt and running are respectively $n_t=-3/(2N^2_*)$, $d n_t/d\ln k \simeq -{3}/{N^3_*}$.

After the end of the inflationary expansion the inflaton is a homogeneous condensate of scalar gravitons. The scalaron universally interacts with other elementary particles only with gravitational strength and the inflaton perturbative decay process can be computed. 
The lifetime of the scalaron is rather long and $\varphi$ decays after it has oscillated excessively many times about the minimum of its potential. The universe during scalaron oscillation phase evolves as a pressureless matter dominated phase and the effective value of the equation of state during reheating is to good approximation zero, $\bar{w}_\text{rh}=0$, \cite{Takeda:2014qma}. Thus the $\Delta N_\text{rh}$ given by the expression (\ref{DNb}) reads
\begin{equation}
\left. \Delta N_\text{rh}\right|_{R^2} =  \frac{1}{12}\ln\left( \frac{\rho_\text{end}}{\rho_\text{rh}}\right)\,.
\end{equation}
The energy density of the inflaton at the end of inflation 
is found to be $\rho_\text{end} \simeq (3/2) V_{R^2}(\varphi_\text{end}) \simeq 3.3 \times 10^{-11}M^4_\text{Pl}$. The energy density at the end of reheating, $\rho_\text{rh}=(\pi^2/30)g_{*\text{rh}}T^4_\text{rh}$, 
is determined by the reheating temperature $T_\text{rh}$ and the number of the degrees of freedom $g_*(T_\text{rh})\equiv g_{*\text{rh}}$. In total, for the $R^2$ inflation the expression (\ref{N*}) is recast into
\begin{equation}
\left. N_*\right|_{R^2}=55.9+\frac14 \ln \epsilon_* +\frac14 \ln \frac {V_*}{\rho_\text{end}}+\frac{1}{12}\ln\left(\frac{g_{*\text{rh}}}{100}\right) +\frac13 \ln\left(\frac{T_\text{rh}}{10^{9}\,\text{GeV}}\right) - \Delta N^\text{}_X\,.
\end{equation}
The reheating temperature is estimated by equating $\Gamma_\text{inf}=H$, where $\Gamma_\text{inf}\equiv \Gamma_\varphi$ is the decay rate of the scalar graviton,
\begin{equation} \label{TR1}
\left. T_\text{rh} \, \right|_{R^2} = \left(\frac{\pi^2}{90}g_{*\text{rh}}\right)^{-1/4} \sqrt{\Gamma_\text{inf} M_\text{Pl}}\,\sim \,10^{9} \,\text{GeV}  \, \left(\frac{100}{g_{*\text{rh}}} \right)^{1/4}\,.
\end{equation}
Assuming only Standard Model degrees of freedom, at that energy scales it is $g_{*\text{rh}}=106.75$, thus $T_\text{rh} \sim 10^9$ GeV.
For the $R^2$ we get for the first slow roll parameter $\epsilon_*=(3/4)/N^2_*$ thus $1/4 \ln\epsilon_*=-2.1+1/2\ln(54/N_*)$. In addition the $R^2$ plateau potential changes very slowly with the $\varphi$ value and for $N_*= 45-60$ it is $1/4 \ln({V_*}/{\rho_\text{end}})\approx 0.2$, hence
\begin{equation}
\left.\left. N_*\right|_{R^2}= N^{(\text{th})}\right|_{R^2} -  \Delta N^\text{}_X\, = 54 - \Delta N^\text{}_X\, .
\end{equation}
In the above equation the logarithmic correction  $1/2\ln(54/N_*)$ has been neglected because its value is less than 0.1 for relevant values of the $N_*$. The {\it thermal} $n_s^{(\text{th})}$ value that the standard Starobinsky $R^2$ inflation model predicts at leading order is found when we substitute the {\it thermal}  e-folds number $N^{(\text{th})}=54$ into the Eq. (\ref{IR1}), that is $n_s^{(\text{th})}=0.963.$ In terms of the e-folds number, the other two slow roll parameters for the Starobinsky model read $\eta_V \simeq  -1/N$ and $\xi_V\simeq 1/N^2$. 
 Since the corrections at second order in slow roll at the scalar tilt will not be 
negligible in the future it is crucial to go to order $1/N^2$.
Also, going at next-to-leading order we could probe $\Delta N_X \sim 1$ changes that could shed light on the pre-BBN cosmic history. For the Starobinsky model the expression (\ref{nsM_next}) reads \cite{Martin:2016iqo}
\begin{equation} \label{nlo}
n_s=1-\frac{\alpha_{R^2}}{N}+\frac{\beta_{R^2}(N)}{N^2}= 1-\frac{2}{N}+\frac{0.81 + 3/2 \ln(N)}{N^2}\,.
\end{equation}
Also, going to order $1/N^3$ the tensor-to-scalar ratio and running read
\begin{equation} \label{raS}
r=\frac{12}{N^2}-\frac{18}{N^3}(2.1+\ln N)\,\quad \text{and} \quad \alpha_s=-\frac{-2}{N^2}+\frac{1}{N^3}(-0.68+3\ln N)\,.
\end{equation}
Plugging  $N^{(\text{th})}=54$ in Eq.(\ref{nlo}) the {\it thermal} scalar tilt value is obtained
\begin{equation}
\left. n_s^{(\text{th})}\right|_{R^2}=0.965\,,
\end{equation}
that is $2\permil$ larger than the leading order prediction. We also take at next-to-leading order
\begin{equation}
\left. \left. r^\text{(th)}\right|_{R^2}=0.0034\,\quad \text{and} \quad \alpha^\text{(th)}_s\right|_{R^2}=-0.037\,.
\end{equation}
Note that the $r$ value is $17\%$ smaller than the value obtained at leading order. 
Furthermore, going to accuracy level $1/N^3$ the $r=r(n_s)$ relation reads
\begin{equation} \label{rn3}
r-3(1-n_s)^2+\frac{23}{4}(1-n_s)^3=0\,.
\end{equation}
The Eq. (\ref{rn3}) was obtained from the expressions $n_s=n_s(\epsilon_V, \eta_V, \xi_V)$ and $r=r(\epsilon_V, \eta_V)$ written up to $1/N^3$ order. In particular for the Starobinsky model it is $n_s-1=2\eta_V-(19/6)\eta^2_V - 2C\eta_V^2+{\cal O}(\eta_V^3)$ and $r=12\eta_V^2+(8-24 C)\eta_V^3 +{\cal O}(\eta_V^4)$ where $C\equiv -2+\ln2+ \gamma$, with $\gamma$ the Euler-Mascheroni constant.
\\
\\
If nature is successfully described by the Standard Model of particle physics and the $R^2$ inflation model then the $\Delta N_X$ has to be zero and hence $n_s=n_s^{(\text{th})}$.
Next we review and estimate the expected $n_s$ and $r$ values for the $R^2$ supergravity inflation model.


\subsection{The $R+R^2$ supergravity inflation}

The embedding of the Starobinsky model of inflation in old-minimal supergravity 
in a superspace 
approach 
consists of reproducing the Lagrangian (\ref{star}).
This is achieved by the action \cite{Cecotti:1987sa,Kallosh:2013lkr,Farakos:2013cqa, Ferrara:2013wka, Dalianis:2014aya} 
\begin{equation}
\label{OM}
{\cal L} = -3 M_P^2 \int d^4 \theta \,  E \,   
\left[ 1 -  \frac{4}{m^2} {\cal R} \bar {\cal R}+  \frac{\zeta}{3 m^4} {\cal R}^2 \bar {\cal R}^2  \right] . 
\end{equation} 
Modifications and further properties can be found in  
\cite{Ellis:2013xoa,  Ellis:2013nxa, Ferrara:2013rsa, Turzynski:2014tza, Kamada:2014gma, Ferrara:2014yna, Ketov:2014hya,
 Alexandre:2013nqa, Kounnas:2014gda, Diamandis:2014vxa, Pallis:2015yyc, Diamandis:2015xra, Farakos:2017mwd, Addazi:2017kbx}. We mention that attention should be paid to the full couplings of the inflaton field that may yield a different reheating temeprature in each of these models since not all of them are pure supergravitational.

The old-minimal supergravity multiplet contains the graviton ($e_m^a$), the gravitino ($\tilde{G}=\psi_m^\alpha$), 
and a pair of auxiliary fields: the complex scalar $M$ and the real vector $b_m$. 
Lagrangian (\ref{OM}) when expanded to components yields $R^2$ terms and kinematic terms for the ``auxiliary'' fields $M$ and $b_m$. 
One may work directly with (\ref{OM}) but it is more convenient to turn to the dual description in terms of 
two chiral superfields:  $T$ and $S$ and standard supergravity \cite{Cecotti:1987sa}. 
During inflation the universe undergoes a quasi de Sitter phase which implies that supersymmetry is broken, the 
the mass of the sgoldstino $S$ becomes large and
it can be integrated out \cite{Lindstrom:1979kq, Farakos:2013ih}. In this stage
a non-linear realization of supersymmetry during inflation is possible  \cite{Antoniadis:2014oya, Ferrara:2014kva,Dall'Agata:2014oka, Kallosh:2014hxa}. 
The real component of $T$ is not integrated out due to the non-linear realization and it is the only dynamic degree of freedom during inflation \cite{Kallosh:2013lkr,Farakos:2013cqa,Dalianis:2014aya}.  
Eventually one finds the effective model (\ref{effS}).

The inflationary predictions for the supergravity $R^2$ model are found to be identical to the non-supersymmetric Starobinsky $R^2$ predictions (\ref{IR1}). In addition, the reheating phase is much similar and the inflaton decay rate roughly the same.  Indeed, in the work of \cite{Terada:2014uia} the inflaton decay channels were identified and the branching ratios calculated. 
The total decay rate was parametrized as $\Gamma_\text{sugra-inf}=c' m^3_\Phi/M^2_\text{Pl}$, where $m_\Phi\equiv m_\text{inf}$ and the reheating temperature was estimated to be
\begin{equation} \label{TR2}
\left. T_\text{rh}\right|_{\text{sugra} R^2} =\left(\frac{90}{\pi^2g_*(T_\text{rh})} \right)^{1/4} \sqrt{\Gamma_\text{sugra-inf} M_\text{Pl}} \,\sim \,10^9 \, \text{GeV}\,.
\end{equation}

The fact that the reheating temperature is found to be about the same with that predicted in the non-supersymmetric $R^2$ model (\ref{TR1}) 
means the supergravity and non-supergravity versions of the $R^2$ inflation models are completely degenerate in terms of the inflationary predictions. However, the details of the expansion history of the universe after the decay of the inflaton should break the degeneracy between the supergravity-$R^2$ and gravity $R^2$.  
We can directly apply the analysis and the results of the previous sections by minimally completing the supergravity $R^2$ sector with the MSSM and a basic supersymmetry breaking sector. 
Let us first examine the implications of the supergravity $R^2$ inflation to the abundances of superparticles. 

The $R^2$ supergravity scenario can be distinguished in two basic cases: the ultra high scale supersymmetry breaking $m_{3/2}> m_\Phi$ and the sub-inflation supersymmetry breaking scale $m_\Phi>m_{3/2}$ case.  The fist case is realized when the minimum of the inflationary potential breaks supersymmetry. Particularly in the model of \cite{Dalianis:2014aya}, where  a new class of  R-symmetry violating $R+R^2$ models was considered, it was found that it is possible inflation and supersymmetry breaking to originate from the supercurvature and obtain $m_{3/2} \sim 2 m_\Phi$ 
without invoking any matter superfields. The new properties of these models which distinguish them from the R-symmetric $R^2$ 
supergravity is that at the end of inflation the $S$ field contribution becomes important. 
In such ultra high scale supersymmetry breaking scenarios the superparticles possibly play no r\^ole during the thermal evolution since the reheating temperature may not be sufficient to excite the superpartners of the Standard Model particles. Hence, the $R^2$ and supergravity $R^2$ models with $m_{3/2}>m_\Phi$ may be totally indistinguishable unless gauginos or some moduli fields are much lighter than the gravitino.

In the case that the inflaton field vacuum is supersymmetric an extra field is required to break supersymmetry, and the condition $m_{3/2}<m_\Phi$  is usually satisfied.  The supersymmetry breaking spurion field called $Z$, i.e. the sgoldstino, although it can play the r\^ole of the diluter it overproduces LSPs and an extra scalar that we generically label $X$  has to play the r\^ole of the diluter\footnote{Any late decaying scalar field, e.g stringy moduli, can be the diluter field.}.
Assuming a simple supersymmetry breaking sector, with $W_{\text{SB}}=FZ+W_0$ and $K_{\text{SB}}=|Z|^2-|Z|^4/\Lambda^2$, the gravitino yield due to the direct decay of the inflaton $\Phi$ is calculated to be \cite{Terada:2014uia}
\begin{equation}
Y^\text{inf}_{3/2}=\frac{3T_\text{rh}}{2m_\Phi} \text{Br}^\text{inf}_{3/2}
\end{equation}
with branching ratio 
\begin{equation} \label{br1}
  \text{Br}^\text{inf}_{3/2}\equiv \text{Br}(\Phi  \rightarrow \tilde{G}\tilde{G}) \simeq \frac{1}{48\pi c'} \times 
  \begin{cases} 
   16 \left( \frac{m_{3/2}}{m_\Phi} \right)^2       \,\,  \quad \text{for} \quad\quad m_Z \ll (m_\Phi m_{3/2})^{1/2}      \\      \\ \left(\frac{m_Z}{m_\Phi} \right)^4 \quad\quad \quad \text{for}   \quad (3m_{3/2}m_\Phi)^{1/2} \ll m_Z  \ll m_\Phi
  \end{cases}
\end{equation}
The $c'$ is determined by the dominant decay channel, here the anomaly induced process \cite{Terada:2014uia}. In the 
case that the spurion field is heavier than the inflaton, $m_\Phi\ll m_Z$, and $m_{3/2}<m_\Phi$ the branching ratio maximizes, $\text{Br}^\text{inf}_{3/2} \simeq (48\pi c')^{-1}$. Otherwise, the gravitino yield is calculated from the branching ratio (\ref{br1}) to be
\begin{equation} \label{YI}
 Y^\text{inf}_{3/2} \simeq \left(\frac{90}{\pi^2 g_{*\text{rh}}}\right)^{1/4} \frac{1}{32\pi} \sqrt{\frac{m_\Phi}{c_x M_\text{Pl}}} \times
  \begin{cases}
     16\left(\frac{m_{3/2}}{m_\Phi}\right)^2  \quad \text{for} \quad\quad m_Z \ll (m_\Phi m_{3/2})^{1/2}      \\
		   \left(\frac{m_Z}{m_\Phi}\right)^4      \,\quad\quad \text{for}   \quad (3m_{3/2}m_\Phi)^{1/2} \ll m_Z\ll m_\Phi
  \end{cases}
\end{equation}
For the supergravity $R^2$ inflation the above contribution to the gravitino abundance  is small. 

Apart from direct gravitino production from inflaton decays, gravitinos are produced via the decay of the $Z$ field. The supersymmetry breaking field $Z$  is produced as particles by the decay of inflaton with branching ratio
\begin{equation}
\text{Br}(\Phi \rightarrow ZZ) = \frac{1}{48\pi c'}\frac{m^2_Z}{m^2_\Phi}\,.
\end{equation}
Considering the generic decay channel 
the $Z$ decays dominantly into a pair of gravitinos when $m_{3/2} \ll m_Z <m_\Phi$
with the partial decay rate enhanced by the factor $(m_Z/m_{3/2})^2$,
\begin{equation}\label{gravZdec}
\Gamma(Z  \rightarrow \tilde{G}\tilde{G})= \left(\frac{m_Z}{m_{3/2}} \right)^2 \frac{m^3_Z}{96 \pi M^2_\text{Pl}}\,.
\end{equation}
Thus, the gravitino yield as a decay product of particle $Z$ is found to be  \cite{Terada:2014uia}
\begin{equation} \label{YZ1}
Y^{Z\text{(particle)}}_{3/2} = \frac{2n_Z}{s} \, = \,2\times 2\times \frac{3T^\text{}_\text{rh}}{4 m_\Phi}\, \text{Br}^Z_{3/2} = \frac{m^2_Z T^\text{}_\text{rh}}{16\pi c' m^3_\Phi},
\end{equation}
where $T_\text{rh}$ the reheating temperature after the decay of the inflaton and $\text{Br}^Z_{3/2}$ the branching ratio of the $Z$ into a pair of gravitinos.
In addition to the incoherent $Z$ particles there are the coherent $Z$ modes, produced by the inflationary de-Sitter phase, which may store a significant amount of energy. The precise VEV of $Z$ is rather model dependent. 
The $\tilde{G}$ yield from the decay of the $Z$ condensate can be computed if the initial amplitude of oscillations $z_0$, the $Z$ mass and couplings are known. 
Assuming a $Z$ dominated universe 
it is
\begin{equation} \label{YZ2}
Y^{Z\text{(cond)}}_{3/2}= 2\times\, \frac{3}{4}\frac{T_Z^\text{dec}}{m_Z} \text{Br}^Z_{3/2}\,,
\end{equation}
and gravitinos and the LSPs are generally found to be overabundant. The initial value and zero temperature VEV of the scalar $Z$ field are rather model dependent and it is possible that the $Z$ scalar does not dominate the energy density of the universe. In the analysis of \cite{Terada:2014uia} the scalar $Z$ is trapped near the origin during inflation.  The zero temperature VEV, dictated by the K\"ahler, $K_{\text{SB}}$, and the superpotential, $W_{\text{SB}}$, is $\left\langle z \right\rangle=2\sqrt{3}(m_{3/2}/m_Z)^2 M_\text{Pl}$.  

In the following we assume benchmark sparticle mass patterns and we estimate the corresponding shift in the spectral index and the tensor-to-scalar ratio in order the predicted dark matter density to be in accordance with observations. We assume gravitino and neutralino dark matter scenarios. We generally assume the presence of an extra scalar labeled $X$ that dilutes the LSP abundance at the critical $\Omega_\text{LSP}h^2=0.12$ and sub-critical values. Particular hidden sector details concerning the $X$ dynamics are left unspecified except for the requirement the diluter not to overproduce LSPs at the time of late entropy production. This is achieved is if the branching ratio to LSPs is very suppressed or $m_\text{LSP}<m_X<2 m_\text{LSP}$.


\subsubsection{The shift in the scalar spectral index and the tensor-to-scalar ratio for the Starobinsky $R^2$ inflation}

The diluter $X$ field dominates the energy density of the universe if $T^\text{dom}_X> T_X^\text{dec}$ where
	  \begin{equation} \label{TXdom}
 T_X^\text{dom}  \simeq
  \begin{cases}
  \left(\frac{x_0}{\sqrt{3} M_\text{Pl}}\right)^2 T_\text{rh}\,,          \,\, \, \, \quad\quad\quad\, \text{for}   \quad \text{scalar condensate} \\
  \left(\frac{30}{\pi^2 g_*}V_0 \right)^{1/4} \, ,  \quad\quad \quad\quad\, \text{for}   \quad  \text{thermal inflation}
		\end{cases}
  \end{equation}
the temperature that the energy density stored in the oscillating $X$ field gets over the radiation energy density. The $x_0$ is the initial amplitude of the oscillations in a potential $V(X)=m^2_X X^2$ about the minimum and $V_0$ the vacuum energy of the flaton field $X$ in the case of thermal inflation. The reheating temperature for Starobinsky and supergravity Starobinsky inflation is $T_\text{rh} \sim 10^9$ GeV and the decay temperature of the $X$ field depends on its full interactions. 
	
The non-thermal $X$ field domination induces a shift in the spectral index value $n_s^{(\text{th})}=0.965$ due to a change in the {\it thermal} e-folds number $N^{(\text{th})}=54$. According to the formula (\ref{Dn}) the size of the shift due to a non-thermal phase that lasts $\tilde{N}_X=[(1-3\bar{w}_X)/4]^{-1} \Delta N_X$ e-folds after the inflaton decay for the Starobinsky model is 
\begin{equation} 
\Delta n_s = -6.3 \times 10^{-4} \,\Delta N_X \left[\sum^{2}_{p=0}\left(0.019 \,\Delta N_X \right)^p - 0.053 \right] \,.
\end{equation}
For a scalar condensate domination it is $\Delta N_X=\ln \tilde{D}_X/3$. The $\Delta n_s$  depends on the dilution size $D_X$ plus a  correction $\hat{g}$ due to the change of the number of the effective degrees of freedom at the temperatures  $T^\text{dom}_X$ and  $T_X^\text{dec}$. Keeping only the relevant terms and after analyzing $\ln\tilde{D}_X=\ln D_X+\hat{g}$ the shift in the scalar tilt reads
\begin{equation} \label{DnR2}
	\Delta n_s (D_X, \hat{g}) =-2 \times 10^{-4} \,(\ln D_X+\hat{g}) \left[1+\frac{2}{300} (\ln D_X+\hat{g}) \right]\,,
\end{equation}
where $\hat{g}\equiv \ln [g_*(T^\text{dom}_X)/g_*(T_X^\text{dec})]/4$.

The shift in the tensor-to-scalar ratio is found by expanding the expression (\ref{raS}) for the $r$, 
\begin{equation}
\Delta r = r(N^\text{(th)}-\Delta N_X)-r(N^\text{th}) = 
   24 \, \frac{\Delta N_X}{\left(N^\text{(th)}\right)^3} +36 \frac{\Delta N_X^2}{\left(N^\text{(th)}\right)^4}  +{\cal O}\left( \frac{\Delta N_X}{(N^\text{(th)})^4}\right)\,.
\end{equation}
Substituting $N^\text{(th)}=54$ and $\Delta N_X =\ln\tilde{D}_X/3 = (\ln D_X +\hat{g})/3$  and keeping only the relevant terms, the above expression for the Starobinsky $R^2$ inflation model reads
\begin{equation} \label{DrR2}
\Delta r(D_X, \hat{g})= 3.9 \times 10^{-5} \,\left(\ln D_X +\hat{g}\right)\left[1+ 8.2 \times 10^{-3} (\ln D_X +\hat{g})\right]\,.
\end{equation}
We have verified that the value $r^\text{(th)}+\Delta r(D_X, \hat{g})$, that the Eq. (\ref{DrR2}) yields, agrees with $10^{-4}$ precision with the value one gets from the relation $r=r(n_s)$ given by the Eq. (\ref{rn3}) for $n_s=n_s^\text{(th)}+\Delta n_s(D_X, \hat{g})$.

Regarding the effective degrees of freedom, it is $\hat{g} \lesssim {\cal O}(1)$, hence the change in the number of degrees of freedom requires accuracy at the $n_s$ $(r)$ measurement of the order of $10^{-4}$ ($10^{-5}$) and one can safely neglect the $\hat{g}$ correction in the expressions (\ref{DnR2}) and (\ref{DrR2}) since the expected accuracy of the future CMB probes will be of the order of $10^{-3}$. 
Nevertheless observing that, in principle at least, one can additionally determine the number of the effective degrees of freedom at the thermal plasma from the $(n_s, r)$ precision measurement is certainly important and exciting, see Fig. 6.


\begin{figure} \label{dndr}
\centering
\begin{tabular}{c} 
{} \includegraphics [scale=.72, angle=0]{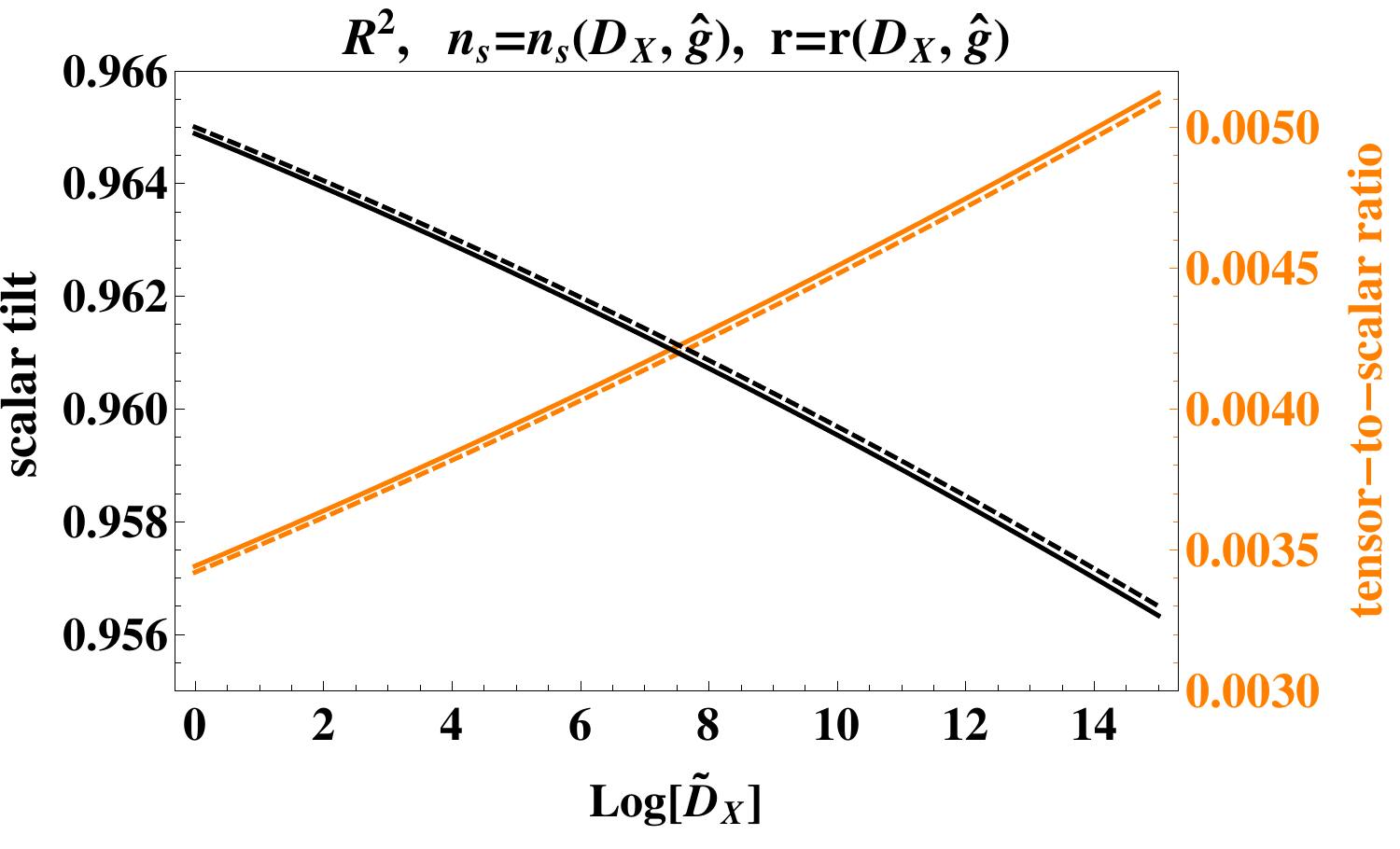} 
\end{tabular}
\caption {\small {The scalar tilt (in black) and tensor-to-scalar ratio (in orange) values when post-inflationary dilution is considered for the Starobinsky $R^2$ inflation model. The solid line corresponds to a change of factor 10 in the number of effective degrees of freedom    in the energy density at the times $T_X^\text{dom}$ and $T_X^\text{dec}$, i.e. $g_*(T_X^\text{dom})=10 \, g_*(T_X^\text{dec})$, and the dashed line corresponds to no change, i.e. $g_*(T_X^\text{dom})=g_*(T_X^\text{dec})$.
}}
\end{figure}


\subsubsection{The $n_s$ and $r$ predictions for particular supersymmetry breaking examples}

In this subsection we explore the impact on $(n_s, r)$ observables of the two base case dark matter scenarios of supersymmetry,  the gravitino and the neutralino, when the initial conditions for the hot Big Bang are set by the supergravity Starobinsky inflation. 
We consider both thermal and non-thermal dark matter production from the hot plasma and scalar decays.
We examine different and illustrative supersymmetry breaking schemes and we quantify how the expected values for the inflationary observables change due to a non-thermal post-reheating phase dictated by the universal constraint $\Omega_\text{LSP}h^2 \leq 0.12$. We mention that this analysis, that probes cosmologically a BSM scheme,  can be applied to any other inflationary model after the appropriate adjustments regarding the reheating phase, the reheating temperature and the inflaton field branching ratios.
\\
\\
\underline{\textbf{{\itshape Example I: Gravitino Dark Matter.} }}
The gravitino is the LSP if the supersymmetry breaking is mediated more efficiently to the MSSM than to the supergravity sector. The standard paradigm is the gauge mediation scenario \cite{Giudice:1998bp}. In such a scenario the supersymmetry breaking $Z$ field decays dominantly into MSSM fields with non-gravitational interactions. Following realistic models \cite{Ibe:2006rc, Hamaguchi:2009hy, Fukushima:2012ra}, it is the imaginary part of the $Z$ field that decays last and the dominant channel is onto a pair of gauginos, in particular binos, with the decay temperature  given by
\begin{equation} \label{Thh}
T^\text{dec}_Z \simeq 760 \text{MeV} \left(\frac{15}{g_*}\right)^{1/4} 
\left(\frac{m_Z}{\text{TeV}}\right)^{1/2} \left(\frac{\text{GeV}}{m_{3/2}} \right) 
\left(\frac{m_{\tilde{g}}}{\text{TeV}}\right)^2 
\left(1-4\frac{m^2_{\tilde{g}}}{m^2_Z} \right)^{1/4}\,.
\end{equation}
The LSP gravitinos are produced from thermal scatterings and decays in the plasma and from the non-thermal decay of the $Z$ scalar field and the inflaton. 
The inflaton contribution to the gravitino abundance is given by Eq. (\ref{YI}) and in general is found to be subleading in $R^2$ inflation.
The decay rate of the $Z$ scalar to gravitinos is given by Eq. (\ref{gravZdec}). 
If the $Z$ decay produces late entropy then the gravitinos from the $Z$ decay, with branching ratio $\text{Br}^Z_{3/2}$ will be part of the dark matter in the universe with yield $Y^Z_{3/2} \sim (3/2) \text{Br}^Z_{3/2} T^\text{dec}_Z/m_Z $ and relic density parameter \cite{Hamaguchi:2009hy} 
\begin{equation} \label{Zgrav}
\frac{\Omega^Z_{3/2}}{0.12\, h^{-2}}  \sim  \left(\frac{15}{g_*} \right)^{1/4} \left( \frac{m_Z}{\text{TeV}}\right)^{7/2} 
\left(\frac{\text{TeV}}{m_{\tilde{g}}} \right)^2 
\left(1-4\frac{m^2_{\tilde{g}}}{m^2_Z} \right)^{-1/4} \,,
\end{equation}
where $m_{\tilde{g}}$ the mass of the bino. We mention that it is also possible that the spurion field does not dominate the energy density due to thermal effects  \cite{Dalianis:2010pq, Dalianis:2013pya}.

Before proceeding with the survey of particular examples, let us mention that the gravitino relic density parameter violates the observational bound unless the sparticles lay in the TeV and sub-TeV scale. Another scalar field $X$ is required to dilute the thermally produced gravitinos and the energy stored in the oscillations of the supersymmetry breaking field, in case of $Z$ domination. 
In order the precise dilution size to be determined the knowledge of the $m_Z$, $m_{3/2}$ and the MSSM mass pattern is necessary.

Let us now consider four benchmark mass patterns for the supersymmetry breaking sector plus the MSSM, with  different sizes of supersymmetry breaking scale. We also consider the presence of messenger fields and the diluter $X$ field necessary to decrease the LSP relic density and which dominantly decays to visible sector fields and not to gravitinos.

\begin{enumerate}

  \item   $\boldsymbol{m_{3/2}\simeq 10^2 \,\text{\textbf{GeV}},  \, m_{\tilde{f}} \sim m_{\tilde{g}}\sim m_Z \simeq 10^4 \, \text{\textbf{GeV} and} \, M_\text{mess} \simeq 10^8  \,\text{\textbf{GeV}}}$. 
	The messengers get thermalized since $M_\text{mess}<T_\text{rh}$ and the scalar spurion  field $Z$ follows the finite temperature minimum without sizable oscillations and hence does {\it not} dominate the energy density \cite{Dalianis:2010pq, Dalianis:2013pya}. We also assume that the messenger coupling is small enough, $\lambda_\text{mess}\ll 1$, so that the gravitinos do not get thermalized. 
The gravitinos produced from {\it scatterings of thermalized messengers} would have a relic density parameter $\Omega^<_{3/2}h^2 \sim 10^4$, see below Eq. (\ref{Oth1}).
The $\Omega_{3/2} h^2 \leq 0.12$ bound implies that the thermally produced gravitinos are sufficiently diluted if $D_X \gtrsim 10^4$. This dilution can be caused by scalar condensate $X$ with $D_X\simeq T^\text{dom}_X/T^\text{dec}_X$. The shift in the spectral index and tensor-to-scalar ratio are respectively $|\Delta n_s| \gtrsim 2\times 10^{-3}$ and $\Delta r \gtrsim 4 \times 10^{-4}$.

		\item 
		$\boldsymbol{m_{3/2}\simeq 10^3 \,\text{\textbf{GeV}},  \, m_{\tilde{f}} \sim m_{\tilde{g}}\sim m_Z > m_{3/2} \, \text{and} \, M_\text{mess} < T_\text{rh}}$. 
		Messengers get thermalized and $Z$ does {\it not} dominate the energy density of the universe. The gravitinos obtain a {\it thermal equilibrium} abundance due to interactions with the {\it thermalized messengers} \cite{Dalianis:2013pya} and their relic density would be $\Omega^<_{3/2}h^2 \sim 10^{10}$, see Eq. (\ref{Oth1}).
The $\Omega_{3/2} h^2 \leq 0.12$ bound implies that the thermally produced gravitinos are sufficiently diluted if $D_X \gtrsim 10^{10}$. The diluter can be either a flaton field that causes thermal inflation or a scalar condensate. In the later case the $X$ field dominates the energy density of the universe shortly after the reheating in order such a dilution size to be realized. 
The shift in the spectral index and tensor-to-scalar ratio are respectively $|\Delta n_s| \gtrsim 5\times 10^{-3}$ and $\Delta r \gtrsim 10 \times 10^{-4}$.

	\item 
	$\boldsymbol{m_{3/2}\simeq 10^4 \,\text{\textbf{GeV}},  \,m_{\tilde{g}}\simeq  10^5 \,\text{\textbf{GeV}}\,, m_{\tilde{f}} \sim m_Z \simeq 10^6 \, \text{\textbf{GeV} and} \, M_\text{mess}>T_\text{rh}}$. 
	The $Z$ field does not receive thermal corrections because the messengers are not thermalized. The $Z$ scalar oscillations generally have a large enough amplitude and $Z$ {\it does} dominate the energy density of the universe. Equations (\ref{Thh}) and (\ref{Zgrav}) say that the spurion $Z$ decays at $T_Z^\text{dec} \simeq 1$ GeV and produces {\it non-thermally} gravitinos that exceed about $10^{6.5}$ times the observational bound.  
	In order the $Z$ condensate to get diluted the $X$ field has to be a flaton and cause thermal inflation. 
	In this case, the shift in the spectral index and tensor-to-scalar ratio are respectively $|\Delta n_s| \gtrsim 3\times 10^{-3}$ and $\Delta r \gtrsim 7\times 10^{-4}$.

	\item $\boldsymbol{m_{3/2}= \text{\textbf{few}}    \,\text{\textbf{GeV}},  \, m_{\tilde{g}}\sim m_{\tilde{f}} \sim m_Z=  \text{\textbf{few}} \,\text{\textbf{TeV}}}$. 
	There are scenarios in the literature that reconcile gravitino cosmology with high reheating temperatures \cite{Dalianis:2011ic, Dalianis:2013pya, Badziak:2015dyc, Co:2017pyf} and generally assume {\it non-minimal} features for the {\it hidden sector}.
For example when the messengers masses lay in the range $M_\text{mess}\lesssim 10^6$ GeV and the goldstino does not reside in a single chiral superfield  \cite{Dalianis:2013pya}, or when the messenger coupling is controlled by the VEV of another field \cite{Badziak:2015dyc} it is possible that gravitinos have the right abundance. 
	These supersymmetry breaking schemes do not require dilution and predict $\Delta n_s=0$ and $\Delta r=0$. We mention that these scenarios, in their original versions, work better when supersymmetry is broken about the TeV scale. Features of these scenarios are currently tested by the LHC experiments.
	
\end{enumerate}

\vspace{3mm}

\begin{table}
\begin{center}
 \begin{tabular}{|c| c c c |c |c| c c c|c|} 
 \hline
 \# & $\boldsymbol{m_Z}$ & $\boldsymbol{m_{\tilde{g}}}$ & $\boldsymbol{m_{\tilde{f}}}$  & $\boldsymbol{m_{3/2}}$ (\textbf{LSP}) & $\boldsymbol{D_{X^{}}}$ &  $ \boldsymbol{N_*}$ & $\boldsymbol{n_s}$ & $\boldsymbol{r}$ & Origin \\ [0.5ex] 
 \hline\hline
 1 & $10^4$ & $ 10^4$ & $10^4$ & $10^2$ & $10^4|_\text{min}$  &  $  51|_\text{max} $   &   $\boldsymbol{0.963|_\text{max}}$ &  $\boldsymbol{0.0038|_\text{min}}$ & Th \\ 
 \hline
 2 & $10^4$ & $10^4$ & $10^5$ & $10^3$ & $10^{10}|_\text{min}$ & $  46|_\text{max}$ & $\boldsymbol{0.960|_\text{max}}$ & $\boldsymbol{0.0044|_\text{min}}$ & Th \\ 
 \hline
 3 & $10^6$ & $10^5$ & $10^6$ & $10^4$ & $10^6|_\text{min}$ & $  49|_\text{max} $ & $ \boldsymbol{0.962|_\text{max}}$ & $\boldsymbol{0.0041|_\text{min}}$ & Non-th\\
  \hline
	4 & $10^3$ & $10^3$ & $10^4$ & $10$ & $1$ & $  54 $ & $ \boldsymbol{0.965}$ & $\boldsymbol{0.0034}$ & Th\\
	\hline
 \end{tabular}
\caption{The $n_s$ and $r$ prediction for gravitino LSP and a gauge mediation scheme for the $R^2$ supergravity model. In the cases $\#$ 1, 2 and 4 the gravitinos are produced from thermal scatterings of messengers and MSSM fields while in the case $\#$ 3 from the non-thermal decay of the supersymmetry breaking $Z$ field. In cases $\#$ 1, 2 and 3 dilution is required to decrease the LSP abundance below the observational bound. In the case $\#$ 4 non-minimal hidden sector features have been assumed.
The masses are in GeV units.} 
  \label{Tgrav}
\end{center}
\end{table}
The above benchmark examples for the gravitino dark matter scenario are synopsized in the table \ref{Tgrav} and Fig. 8.
\\
\\
\underline{\textbf{{\itshape Example $II$: Neutralino Dark Matter.}	}}
For gravity or anomaly mediation of supersymmetry breaking the gravitino mass is naturally heavier than the neutralinos. The gravitino decay populates the universe with neutralinos.
Here we assume the gravitino mass to be above $10^5$ GeV not to spoil BBN predictions at the time of decay. 
The gravitinos are produced non-thermally by the decay of the inflaton,  see Eq. (\ref{YI}), which generally accounts for a subleading contribution in the framework of $R^2$ supergravity inflation, and by the decay of the supersymmetry breaking scalar field $Z$. Contrary to the GMSB case the $Z$ scalar oscillations are not thermally damped and generally the $Z$ produces late entropy if displaced from the zero temperature minimum.
The temperature that the $Z$ field decays is estimated by considering the various partial decay rates. The dominant decay channel is into a pair of gravitinos, when  $m_Z \gg m_{3/2}$, and the total decay rate yields the decay temperature 
	\begin{equation}
	T_Z^\text{dec} \simeq 4 \times 10^9 \text{GeV} \left(\frac{m_Z}{10^8\text{GeV}}\right)^{5/2} \left(\frac{\text{GeV}}{m_{3/2}} \right)\,.
	\end{equation}
If the $Z$ field oscillations dilute the thermal plasma then the gravitinos coming from the $Z$ decay are the leading source of dark matter neutralinos at the gravitino decay temperature $T^\text{dec}_{3/2}$.
The neutralinos are generally found to be overabundant when supersymmetry breaks at energies beyond the TeV scale and dilution is required. Hence we assume the presence of a diluter field $X$ that decreases the LSP relic density via late entropy production.
 We mention that according to the general constraint (\ref{c3}) the neutralinos with mass $m_{\tilde{\chi}^0}>10^7$ GeV are impossible to get diluted by the oscillations of the $X$ scalar and thermal inflation is required.

Let us now consider benchmark mass patterns for the supersymmetry breaking sector plus the MSSM, characterized mainly by split and quasi-natural sparticle mass spectrum.

\begin{table}
\begin{center}
 \begin{tabular}{|c| c c c |c | c| c c c | c|} 
 \hline
 \# & $\boldsymbol{m_Z}$ & $\boldsymbol{m_{3/2}}$ & $\boldsymbol{m_{\tilde{f}}}$  & $\boldsymbol{m_{\tilde{\chi}^0}}$ (\textbf{LSP}) &  $\boldsymbol{D_{(X)}}$  & $ \boldsymbol{N_*}$ & $\boldsymbol{n_s}$ & $\boldsymbol{r}$  & Origin \\ [0.5ex] 
 \hline\hline
 1 & $10^7$ &  $10^6$ & $10^6$ &  $10^3$ & $10^2|_\text{min}$ & $ 52|_\text{max} $ & $\boldsymbol{0.964|_\text{max}}$ & $\boldsymbol{0.0036|_\text{min}}$ & Non-th\\
 \hline
 2 & $10^9$ & $ 10^8$ & $10^8$ &  $10^3$ & $10^2|_\text{min}$ & $ 52|_\text{max}$ & $\boldsymbol{0.964|_\text{max}}$ & $\boldsymbol{0.0036|_\text{min}}$ & Th \\ 
 \hline
  3 & $10^8$ & $10^7$ & $10^7$ &  $10^5$ & $10^8|_\text{min}$ &$ 48|_\text{max}$ & $\boldsymbol{0.961|_\text{max}}$ & $\boldsymbol{0.0042|_\text{min}}$ & Non-th \\ [1ex] 
 \hline
4 & $10^5$ & $10^5$ & $10^5$ &  $10^3$ & $1$ &$ 54$ & $\boldsymbol{0.965}$ & $\boldsymbol{0.0034}$ & Th \\ 
 \hline
\end{tabular}
\caption{The $n_s$ and $r$ prediction for neutralino LSP and anomaly/gravity mediation scheme for the $R^2$ supergravity model. In the case $\#$ 1 the neutralino annihilate after the decay of gravitinos, while in case $\#$ 2 neutralinos acquire a thermal abundance. In the case $\#$ 3 the neutralinos from the gravitino decay are overabundant and a diluter $X$ is required. The case $\#$ 4 is the standard thermal WIMP scenario. The masses are in GeV units. } 
  \label{Tneu}
\end{center}
\end{table}

\begin{enumerate}
	
	\item 
	$\boldsymbol{m_{\tilde{\chi}^0}\lesssim 10^3}$ \textbf{GeV,} 
	$\boldsymbol{m_{3/2} \sim m_{\tilde{f}} \simeq 10^6 \, \text{\textbf{GeV}}, m_Z\simeq  10^7}$ \textbf{GeV}.
Here we assume the {\it annihilation scenario} where the neutralino has an annihilation cross section few orders of magnitude higher that the conventional value. The universe is generally dominated by the $Z$ scalar that decays to gravitinos at the temperature $T^\text{dec}_Z\sim 12$ GeV. In turn, the gravitinos produced from the $Z$ decay dominate the energy density and decay at $T^\text{dec}_{3/2}\sim 0.2$ GeV producing neutralinos that annihilate rapidly and acquire a relic density $\Omega^\text{(ann)}_{\tilde{\chi}^0}  = \Omega^\text{(th)}_{\tilde{\chi}^0}   \left ({T^\text{f.o.}_{\tilde{\chi}^0}}/{T^\text{dec}_{3/2}}\right)$, see section 3.2. The resulting LSP abundance can fit the observed value and here the r\^ole of the diluter is played by the $Z$ field and the gravitinos. 
We note that if $m_{\tilde{\chi}^0}>$ TeV then a diluter scalar $X$ is required.
It is $D \gtrsim 10^2$ but the dilution size due to $Z$ oscillations can be many orders of magnitude larger. This scenario is currently tested and constrained by the LHC and indirect detection experiments  \cite{Easther:2013nga}. 
This minimum value of the dilution magnitude yields $|\Delta n_s| \gtrsim 1 \times 10^{-3}$ and $\Delta r \gtrsim 2 \times 10^{-4}$.

  \item 
	$\boldsymbol{m_{\tilde{\chi}^0}\simeq 10^3}$ \textbf{GeV,} 
	$\boldsymbol{m_{3/2} \sim m_{\tilde{f}} \simeq 10^8 \, \text{\textbf{GeV}}, m_Z \simeq 10^9}$ \textbf{GeV}.
In this example we assume that the $T^\text{dec}_{3/2}>T^\text{f.o.}_{\tilde{\chi}^0}$ and {\it neutralinos thermalize} after the decay of gravitinos.
A $Z$ dominated early universe  becomes in turn gravitino dominated at $T^\text{dec}_Z \sim 10^4$ GeV. For TeV and sub-TeV scale neutralinos the observational bound $\Omega_\text{DM}h^2 \sim 0.12$ can be satisfied, see Eq. (\ref{neuAb}). Here again the r\^ole of the diluter is played by the $Z$ field and the gravitinos and it is  $D \gtrsim 10^2$, but it can many orders of magnitude larger. This scenario is currently tested by LHC and direct detection experiments.
This dilution magnitude induces a shift in the spectral index and tensor-to-scalar ratio respectively at least of size $|\Delta n_s| \gtrsim 1\times 10^{-3}$ and $\Delta r \gtrsim 2 \times 10^{-4}$.

\item
$\boldsymbol{m_{\tilde{\chi}^0}\simeq 10^5}$ \textbf{GeV,} 
	$\boldsymbol{m_{3/2} \sim m_{\tilde{f}} \simeq 10^7 \, \text{\textbf{GeV}}, m_Z\simeq 10^8}$ \textbf{GeV}.
In this example the {\it neutralinos are produced from the gravitino decay} and they are {\it out} of chemical and kinetic equilibrium. The LSP relic density, given by Eq. (\ref{GD}), is $\Omega^<_{\tilde{\chi}^0} h^2 \sim 10^8$. The LSP abundance has to be decreased eight orders of magnitude down and this is possible only if the gravitinos and the $Z$ scalar condensate are sufficiently diluted. Thermal inflation is required with $D_X \gtrsim 10^8$. This dilution magnitude induces a shift in the spectral index and tensor-to-scalar ratio respectively at least of size $|\Delta n_s| \gtrsim 4\times 10^{-3}$ and $\Delta r \gtrsim 8 \times 10^{-4}$. This is a phenomenologically viable example not constrained by terrestrial experiments.

\item $\boldsymbol{m_{\tilde{\chi}^0}\simeq 10^3}$ \textbf{GeV,} 
	$\boldsymbol{m_{3/2} \sim m_{\tilde{f}} \sim m_Z> 10^5}$ \textbf{GeV}. 
As a last example we consider the conventional {\it thermal WIMP} scenario assuming that the $Z$ scalar field is not displaced from the zero temperature minimum and never dominates the energy density of the universe, hence {\it no} non-thermal phase is required (although a non-thermal phase before $T^\text{f.o.}_{\tilde{\chi}^0}$ is not ruled out in general). 
In this scenario it is $\Delta n_s=0$ and $\Delta r=0$.
This supersymmetry breaking scheme is currently tested at LHC, direct and indirect detection experiments.
\end{enumerate}
The above benchmark examples for the neutralino dark matter scenario are synopsized in the table \ref{Tneu} and Fig. 8.
\\
\\
Let us finally note that the LSP particles produced from the gravitino and $X$ decay are warmer than the LSPs produced from thermal scatterings and this changes the free streaming length of the LSP dark matter, which has the effect of potentially washing out small scale cosmological perturbations, see e.g. \cite{Kelso:2013paa, Allahverdi:2014bva}. This is a very interesting possibility that could provide further constraints to these scenarios, though the mass scales and lifetimes considered here yield free streaming lengths that are not in conflict with the Lyman-$\alpha$ forest observations \cite{Viel:2005qj}.

\subsection{Distinguishing the $R^2$ and the $R^2$ supergravity inflationary models}

The supersymmetric and non-supersymmetric $R^2$ inflation models predict the same reheating temperature, $T_\text{rh} \sim 10^9$ GeV and the same expressions for the $n_s=n_s(N)$ and $r=r(N)$.
However, the degeneracy between the two models that appears during the accelerating and the reheating stage breaks after the inflaton decay\footnote{The present comparison of the $R^2$ and supergravity $R^2$ inflation can be viewed as complementary to the analysis of \cite{Dalianis:2015fpa} that focused on the initial conditions of the two models.}. In the case of supergravity $R^2$ inflation, if $\tilde{m}< T_\text{rh}$, sparticles will be constituents of 
the thermalized plasma of the reheated universe. In addition to thermal processes, the presence of the supergravitational inflaton and the supersymmetry breaking field produce a significant number of gravitino particles after inflation, as the expressions (\ref{YI}), (\ref{YZ1}) and (\ref{YZ2}) make manifest.

The BBN and the $\Omega_\text{DM}h^2=0.12$ constraints imply that the thermal cosmic history is influenced by the change of the supersymmetry breaking pattern. The LSP is found to be overabundant in the greatest part of the MSSM parameter space and it receives further contributions when the supersymmetry breaking field is taken into account \cite{Terada:2014uia} 
for both gravitino and neutralino LSP scenarios. 
Therefore, $R^2$ supergravity inflation is compatible with the cosmological observations only if the thermal history of the universe is not perpetual from $T_\text{rh}\sim 10^9$ GeV until $T_\text{BBN}\sim 1$ MeV. 
A non-thermal phase that dilutes the supersymmetric thermal relics and potentially supplements the universe with dark matter particles 
can fully reconcile the $R^2$ supergravity inflation model with observations. 
The required dilution generally increases with increasing the sparticle masses. Henceforth, we conclude that the degeneracy breaking of the inflationary predictions between the $R^2$ and supergravity $R^2$  models depends on the energy scale and the pattern of supersymmetry breaking, see Fig. 7 and 8.

\begin{figure} \label{f01}
\centering
\includegraphics [scale=.65, angle=0]{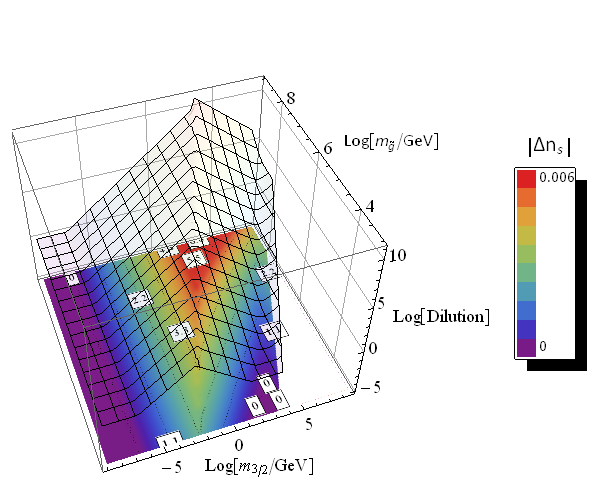} 
\caption{\small{ This 
is a compound plot consisting of 3D graph and a density-contour plot. The 3D graph shows the decadic logarithm of the {\itshape required} dilution magnitude as a function of the gravitino LSP and gaugino-sfermion masses with $m_{\tilde{f}}=m_{\tilde{g}}$ for a reheating temperature $T_\text{rh}=10^9$ GeV. The dilution is calculated by requiring the $\Omega_{3/2}h^2$ not to exceed the observational bounds. 
The density-contour plot demonstrates the change in the $n_{s}$ value, magnified 1000 times on the contour labels, for inflationary models that predict a reheating temperature $T_\text{rh}= 10^9$ GeV.  
The information that one extracts from this graph is that supersymmetric models (e.g. quasi-natural, split, high scale) can be compatible with the CMB data only for particular values for the scalar tilt $n_s$. 
}}
\end{figure}

The fact that the $R^2$ supergravity automatically alters the details of the thermal history and possibly the expansion history of the universe compared to the simple $R^2$ case, where  sparticles and supersymmetry breaking fields are absent, allows the discrimination between the two inflationary models. 
Considering only the MSSM degrees of freedom as the less model dependent and conservative analysis, 
the conditions (A) and (B) of section 4, when true, imply that the supergravity $R^2$ inflationary model predicts
\begin{equation} \label{discr}
 n_s(k_*) < 0.965 \quad\quad \text{and} \quad\quad r_*>0.0034 \,,
\end{equation}
and $r=3(1-n_s)^2-23/4(1-n_s)^3$, which is the characteristic $r=r(n_s)$ relation for the Starobinsky $R^2$ inflationary model, see Fig. 8.  The $n_s=0.965$  and $r=0.0034$ are the reference thermal values.  
A knowledge of the details of the supersymmetry breaking sector would allow us to accurately predict the ($n_s, r$) values.
From a different point of view, the precise measurement of the ($n_s, r$) observables could indicate cosmologically viable supersymmetry breaking patterns. Although it may not be possible to specify the identity of the dark matter, see the proximity of the spots on the ($n_s, r$) contour in Fig. 8, it is possible to  constrain significantly and even rule out a great part of the supersymmetry breaking parameter space.

Let us also mention that a similar result to (\ref{discr}) can be obtained if one simply assumes the presence of extra scalars that dominate the energy density of the early Universe.
For example, for a supergravity $R^2$ inflation model, a gravitational modulated reheating was assumed \cite{Watanabe:2013ppx} and non-Gaussianity  was additionally predicted.
 In the present work the postulation of a non-thermal phase has been motivated by the {\it general requirement to fit the universal constraint} $\Omega_\text{LSP}h^2\leq 0.12$ that in turn implies the result (\ref{discr}).
Last but not least, we emphasize again that the precise measurement of the ($n_s, r$) cannot "prove" or disprove the existence of supersymmetry. It will only indicate the presence of  extra scalar degrees of freedom, that supersymmetry or any other BSM scenario will be challenged to explain.


\begin{figure} \label{prism}
\centering
\includegraphics [scale=1, angle=0]{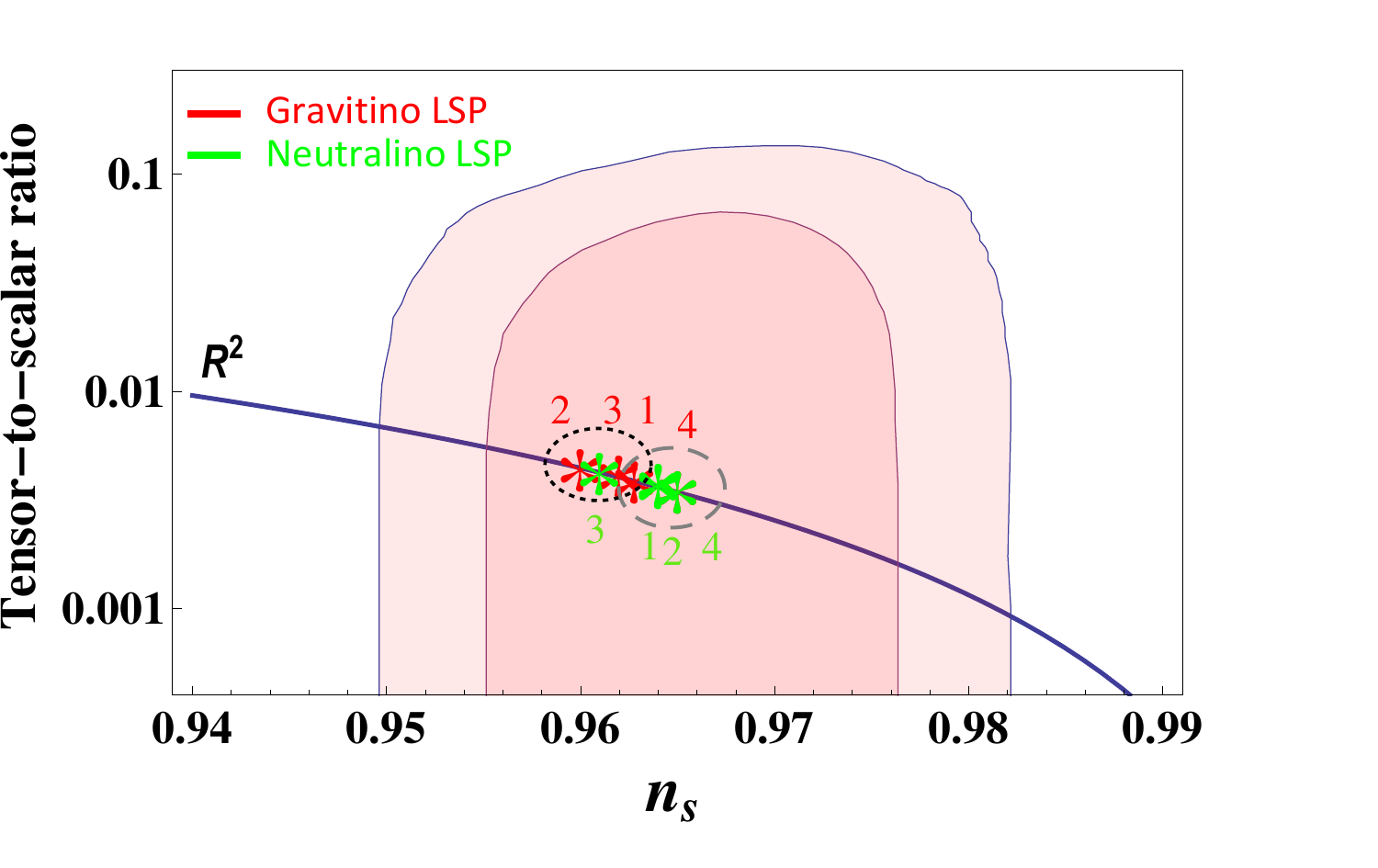} 
\caption {\small {Constraints on the ($n_s, r$) contour plane from Planck-2015 in the pink, and the schematic illustration of $2\sigma$ forecast constraints from a future CMB probe with sensitivity $\delta n_s \sim 10^{-3}$ and $\delta r \sim 10^{-3}$ depicted with the dotted and dashed ellipsis. The $R^2$ model is targeted with a fiducial value of $r \sim 4 \times 10^{-3}$. The red asterisks correspond to the predictions of the four benchmark models $(\# 1,2,3,4)$ with {\it gravitino} LSP and the green asterisks to the four benchmark models $(\# 1,2,3,4)$ with {\it neutralino} LSP, as explained in the text and tables \ref{Tgrav} and \ref{Tneu} respectively. 
If the future CMB experimental probes select the area inside the dashed ellipsis then either the $R^2$ or the SUGRA-$R^2$ inflation model is selected {\itshape plus} a roughly continuous thermal phase with reheating temperature, $T_\text{rh}\sim 10^9$ GeV. The selection of the dashed ellipsis area will exclude a large class of supersymmetry models that predict a too large LSP abundance for that reheating temperature. On the contrary, if the dotted ellipsis area is selected then the duration of the thermal phase before the BBN is much limited and extra scalar particles should be present above the TeV scale, hence supporting the SUGRA-$R^2$ model rather than the $R^2$ inflation model  {\it plus "desert"}. 
}}
\end{figure}


\begin{figure} \label{fgraphI}
\centering
\includegraphics [scale=1.2, angle=0]{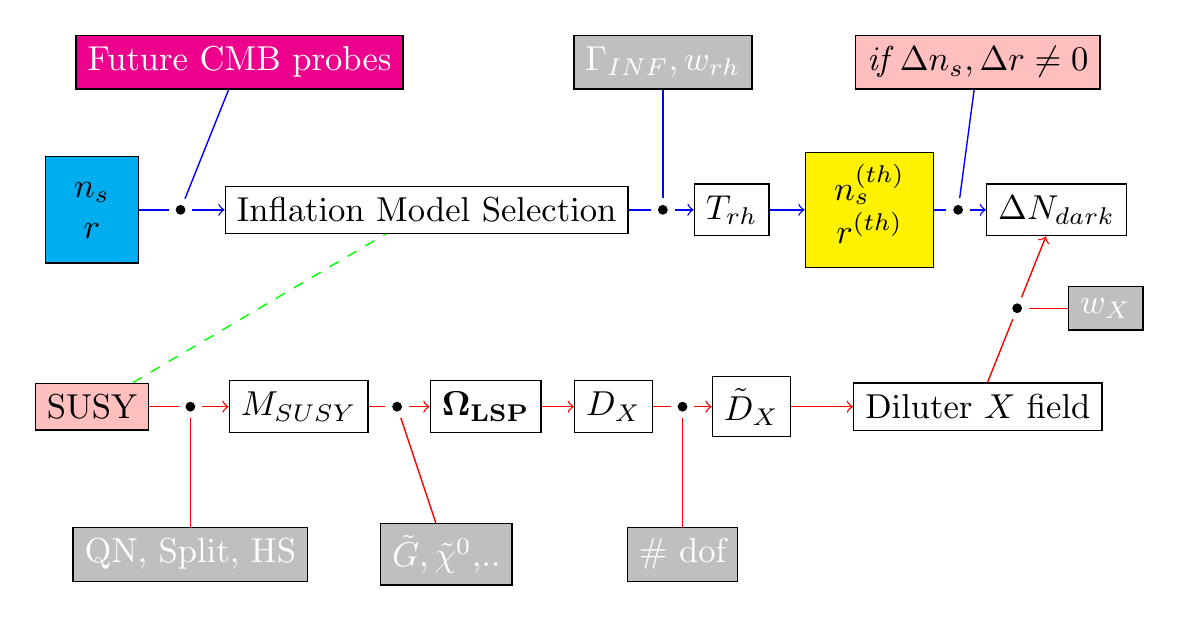} 
\caption{\small{ This graph demonstrates the analysis followed in this work to cosmologically probe BSM scenarios.
}}
\end{figure}

\section{Discussion and Conclusions}

The cosmic energy window from about $1$ MeV up to the inflationary energy scale is shuttered to the current observational probes and the corresponding timescale can be reasonably called a {\it dark} early universe cosmic era. 
Any understanding of the cosmic processes that take place before the BBN will provide us with critical insights into the microphysics that operates at that energy scales.
One significant prospect to contemplate the early dark cosmic era is through the precision measurement of the CMB observables $n_s$ and $r$.
The essential fact is that the $(n_s, r)$ are not strictly scale invariant, hence important information about the background expansion rate and the reheating temperature of the universe can be obtained. 

The inflationary paradigm can be used as a concrete and compelling framework for the theoretical determination of the $n_s$ and $r$ values. However, our ignorance about the reheating process and the subsequent evolution of the universe, encoded in the dependence on $N_*$, is rather strong and will become significant as the accuracy on the observations are expected to be further improved the next decades. 
An inflationary prediction that is independent of $N_*$ is the contour line  $r=r(n_s)$ which can distinguish different inflation models.
Furthermore, if inflation is followed by a continuous thermal phase then a concrete inflation model predicts a specific number for the number of e-folds between the moment the relevant modes exit the horizon and the end of inflation, hence predicts a specific spot on the $r=r(n_s)$ line that corresponds to what we called {\it thermal} values for the e-folds number, scalar tilt and tensor-to-scalar ratio, $N^\text{(th)}$, $n_s^{\text{(th)}}$ and $r^\text{(th)}$  respectively. 

Motivated by the advertised sensitivity of the future CMB probes in this paper
we quantified the effect of a generic primordial non-thermal phase on the spectral index value  (\ref{Dn}).  
The $n_s$ value is possible to have been shifted by the amount $\Delta n_s/ n_s\sim {\cal O}(1-6) \permil$ from the expected thermal value, $n^{(\text{th})}_s$, due to a scalar condensate or a flaton field domination. 
The observation of non zero $\Delta n_s$ and $\Delta r$ along a contour line $r=r(n_s)$ is an indirect observation of a non-thermal phase and connects cosmology to microphysics since it has to be attributed to a BSM scalar field domination.

Moving a step further we applied our general results to study the observational consequences on the CMB of a supersymmetric universe. 
Supersymmetry is one of the most motivated theories that is extensively used to describe the very early universe evolution. Although it lacks any experimental support, it provides an appealing framework that consistently accommodates high energy processes such as inflation and dark matter production. Actually, the fact that the LHC probes only a small part of the vast energy scales up to the Planck mass, while supersymmetry may lay anywhere in between, strongly motivates the systematic cosmological examination of supersymmetric scenarios.  Supersymmetry can be cosmologically manifest if  supersymmetric degrees of freedom get thermally excited or produced non-thermally  during the dark early cosmic era.

The most direct cosmological implication of supersymmetry is that the LSP expected to be stable and hence contributes to the dark matter density. The LSP abundance is the key quantity that we estimate in different classes of supersymmetry breaking schemes and examine how it can be cosmologically reconciled with the observational value $\Omega_\text{DM}h^2=0.12$. 
 We find that a non-thermal phase or low reheating temperatures are generally required if supersymmetry UV completes the Standard Model of particle physics. 
We quantified the effect of the different expansion histories on the $(n_s, r)$ and we broadly related it with the different supersymmetry breaking schemes. In this paper we mostly focused on ultra-TeV scale supersymmetry since  low scale supersymmetry models with thermal WIMPs are in growing conflict with collider data and direct detection experiments. In our analysis we have not assumed that the LSP accounts for the bulk dark matter component in the universe. If it is actually  $\Omega_\text{LSP} h^2\ll 0.12$ then the expected change in the ($n_s, r$) values due to a non-thermal stage becomes greater.

A complete understanding of the pre-BBN thermal phase and the CMB observables requires the knowledge of the initial condition for the thermal Big Bang, which are successfully provided by the inflationary theory. In this work we suggested a unified study of inflation and the subsequent reheating stage. Actually it is often the case that supergravity inflationary models are degenerate, in terms of the inflationary observables, with their the non-supersymmetric versions.
However, the supersymmetric degrees of freedom can be excited either thermally or non-thermally after the end of the inflationary phase. 
For the sake of completeness we considered in this paper the $R^2$ supergravity inflation and we performed a theoretical estimation of the  $(n_s, r)$  observables. 
Our findings point out that the ultra-TeV scale supersymmetry leaves a more clear cosmological imprint on the CMB observables. This fact is particularly exciting because high scale supersymmetric scenarios can be cosmologically falsified while
  the low mass range supersymmetric scenarios are directly tested at the terrestrial colliders.

Undoubtedly any non-trivial cosmological information about the BSM physics is of major importance.  Certainly the results of this cosmological analysis, illustrated in the graph of Fig. 9, cannot discover or disprove supersymmetry. The only concrete cosmological information that we get from the $n_s$ and $r$ observables concerns the expansion rate of the very early universe. 
The identity of the  matter content that controls the cosmic expansion rate cannot be revealed and it is only subject to interpretations. Nonetheless, if the $(n_s, r)$ deviate from their thermal values then new physics exists in high energies.
In this paper we focused on supersymmetry, though any BSM scenario can be analyzed accordingly.
In the event of detection of primordial gravitational waves, that is observation of $r\neq 0$ together with possible features of the tensor power spectrum, then the selection of a particular inflation model is possible. In such a case our analysis has  the power to rule out the BSM desert scenario and indicate possible features of candidate BSM theories, as the Fig. 8 illustrates.

From the theoretical side a more complete analysis should also take into account baryognesis scenarios and the details of 
thermalization process. The generation of the matter-antimatter asymmetry in the universe, seems to have a critical dependence on the temperature, as e.g. the thermal leptogenesis scenario \cite{Fukugita:1986hr} suggests.
Moreover the understanding of several distinct stages in the reheating process that leads to thermalization of the universe  in a radiation dominated phase at some reheating temperature $T_\text{rh}$ is  necessary in order a more accurate value for the equation of state parameter $w_\text{rh}$ and the reheating e-folds number $\tilde{N}_\text{rh}$ to be estimated, see Eq. (\ref{DNb}). A thorough understanding of  the reheating process can also bring out new observables that can further constrain the reheating temperature of the universe, see e.g. \cite{Amin:2014eta} for a review. We should mention here that the oscillatory epoch and the reheating process of the $R^2$ inflation model is well understood, a fact that makes the results obtained in section 5 reliable \cite{Takeda:2014qma}.

From the observational side, future CMB primary anisotropy measurements should play a decisive r\^ole in probing the pre-BBN cosmic era.  Complementary observational programs, such as the direct observation of tensor perturbations, should contribute significantly to this endeavor as well. 
Information on the thermal history after inflation is imprinted in the gravitational wave spectrum in the frequencies corresponding to the reheating energy scales, which can be probed by future space-based laser interferometers such as DECIGO \cite{Kuroyanagi:2014qza}.
Presumably,  the synergy of different cosmological surveys will enable a leap forward in precision cosmology 
giving us, at the same time, access to  the physics that operates beyond the Standard Model of particle physics, at energy scales much higher than can be obtained at CERN.




\vspace*{.5cm}

\section*{Acknowledgments}

\noindent 
We thank Fotis Farakos, Alex Kehagias and Jun'ichi Yokoyama for discussions and comments on the draft. 
The work of ID is supported by the IKY Scholarship Programs for Strengthening Post Doctoral Research, co-financed by the European Social Fund ESF and the Greek government. The work of YW is supported by JSPS Grant-in-Aid for Young Scientists (B) No. 16K17712.

	\end{document}